\documentclass[12pt,a4paper]{article}
\usepackage[compress,sort]{cite}
\usepackage[parsep]{collref}
\usepackage{graphicx} 
\usepackage{amssymb, amsmath}
\usepackage{epstopdf}
\usepackage{fullpage}
\usepackage{microtype}

\usepackage{pstricks}
\usepackage{wrapfig,rotating}
\usepackage{multirow}
\numberwithin{equation}{section}

\usepackage{axodraw4j}
\usepackage{feynmf}\usepackage{bbm}
\usepackage{slashed}\usepackage{enumerate}

\usepackage{pdfsync}
\def\ee{e^{+}e^{-}}
 \def\qbar{\bar{q}}
  
 \def\as{\alpha_{s}} 
 
 \def\b1{\beta_{1}}

\def\d{\mathrm{d}}  
\def\ln{\mathrm{ln}} 
\def\Li{\mathrm{Li}} 
 
\def\kto{k_{t1}} \def\ktt{k_{t2}}
\def\k3{\mathbf{k}} \def\kt{k_{t}}

\def\R{\mathcal{R}}\def\O{\mathcal{O}}
\def\pt{p_{t}} 

\def\uv{\hat{\mathbf{n}}}

\def\CA{C_{A}} \def\CF{C_{F}} 
\def\Nc{N_{c}} \def\TF{T_{R}}
 \def\nf{n_{f}}

\def\outt{\mathrm{out}} \def\inn{\mathrm{in}}
\def\Eo{E_{0}}
\def\deta{\Delta\eta}
\def\so{\sigma_{0}}

\def\tauo{\tau_{E_{0}}}

\def\Rs{R_{s}}\def\Or{\mathcal{O}}
\def\f0{f_{0}}
\def\SCET{\mathrm{SCET}}

\def\akt{\mathrm{anti-k_{t}}}
\def\ca{\mathrm{C-A}}

\begin{document}

\begin{titlepage}
\begin{flushright}
{MAN/HEP/2011/16}
\end{flushright}  
\vskip8mm

\begin{center}
{\Large \bf
Non--global logs and clustering impact on jet mass with a jet veto distribution}
\vspace*{1.5cm}

Kamel Khelifa--Kerfa\footnote{\texttt{Kamel.Khelifa@hep.manchester.ac.uk}}
\\
School of Physics \& Astronomy, University of Manchester,\\
Oxford Road, Manchester, M13 9PL, U.K.

\bigskip
\bigskip

 { \bf Abstract }
\end{center}

\begin{quote}
There has recently been much interest in analytical computations of jet mass
distributions with and without vetos on additional jet
activity~\cite{Ellis:2009wj,Ellis:2010rwa,Banfi:2010pa,Kelley:2011tj,
Hornig:2011tg,Li:2011hy}. An important issue affecting such calculations,
particularly at next--to--leading logarithmic (NLL) accuracy, is that of non--global logarithms as well as logarithms induced by jet definition, as we pointed out in an earlier work~\cite{Banfi:2010pa}. In this paper, we extend our previous calculations by independently deriving the full jet--radius analytical form of non--global logarithms, in the anti--$\kt$ jet algorithm. Employing the small--jet radius approximation, we also compute, at fixed--order, the effect of jet clustering on both $\CF^{2}$ and $\CF\CA$ colour channels. Our findings for the $\CF\CA$ channel confirm earlier analytical calculations of non--global logarithms in soft--collinear effective theory~\cite{Hornig:2011tg}. Moreover, all of our results, as well as those of~\cite{Banfi:2010pa}, are compared to the output of the numerical program \texttt{EVENT2}. We find good agreement between analytical and numerical results both with and without final state clustering.
\end{quote}

\end{titlepage}

\newpage

\tableofcontents

\section{Introduction}\label{sec.intro}

Event and jet shape variables have long served as excellent tools for testing
QCD and improving the understanding of its properties (for a review,
see~\cite{Dasgupta:2003iq}). Event/jet shape distributions have been used to
extract some prominent parameters in QCD including the strong coupling and the
quark--gluon colour ratio~\cite{Beneke:1998ui}. Due to the fact that shape
variables are, by construction, linear in momentum, they exhibit a strong
sensitivity to non--perturbative (NP) effects~\cite{Dasgupta:2003iq,Webber:1994zd}. They have thus been exploited to
gain a better analytical insight into this QCD
domain~\cite{Dasgupta:2003iq,Beneke:2000kc}. Furthermore, jet shapes have been
used not only to study the jet structure of hadronic final states, including jet
multiplicities, jet rates and jet profiles (Ref.~\cite{QCD_collider} and
references therein), but also the \emph{subjet} structure, or substructure, of
the jets themselves (for a recent example, see~\cite{Ellis:2009wj}). The latter
subject has received significant attention in recent years, particularly in the
area of boosted objects  with the aim to separate the decay products of Beyond
Standard Model (BSM) particles from QCD background at LHC (for a review,
see~\cite{Abdesselam:2010pt}).

Although shape variables are, by construction, Infrared and Collinear (IRC)
safe, fixed--order perturbative (PT) calculations break down in regions of phase
space where the shape variable is small. These regions correspond to gluon
emissions that are soft and/or collinear to hard legs and lead to the appearance
of large logs that spoil the PT expansion of the shape
distribution~\cite{QCD_collider} (and references therein). While measured shape
distributions have a peak near small values of the shape variable and then go
to zero, fixed--order analytical distributions diverge. To deal away with these
divergences and successfully reproduce the experimentally--seen behaviour, one
ought to either perform an all--orders resummation of the large logs, matched to fixed--order result, or rely on Monte Carlo event generators. We are
concerned, in the present paper, with the resummation method as it paves the way
for a better understanding of QCD dynamics including the process of multiple
gluon radiation. The general form of resummed distributions for observables that have the property of exponentiation can be cast as~\cite{QCD_collider}
\begin{equation}\label{resum_gen_struct}
 \Sigma(v) = C(\as)\,\exp\left[L\,g_{1}(\as L) + g_{2}(\as L) + \as\,g_{3}(\as L)+ \cdots \right] + D(v)
\end{equation}
where $L=\ln(1/v)$, $C(\as)$ is an expansion in $\as$ with constant coefficients that can be inferred from fixed--order calculations and $D(v)$ collects terms that are proportional to powers of the shape variable $v$. The function $g_{1}$ resums all the leading logs (LL) $\as^{n} L^{n+1}$, while $g_{2}$ resums the next--to--leading logs (NLL) $\as^{n} L^{n}$ and so on.

There are two types of jet shape observables~\footnote{from the point of view of
our calculations in this paper.}: global and non--global~\cite{Dasgupta:2001sh}.
\emph{Global} observables are shape variables that are sufficiently inclusive
over the whole final state phase space. The resummation of such variables, e.g,
thrust, heavy jet mass and broadening, up to NLL accuracy have long been
performed~\cite{Catani:1992ua,Banfi:2001bz}. The resultant resummed distributions were then matched with NLO fixed--order results for a better agreement with measurements over a wide range of values of the shape variable~\cite{Catani:1992ua, Banfi:2010xy}. In the recent past, the NNLL $+$ NLO distribution has been obtained for energy--energy correlation~\cite{deFlorian:2004mp}, as well as NNLL $+$ NNLO~\cite{GehrmannDeRidder:2005cm,GehrmannDeRidder:2007hr} for the thrust distribution~\cite{Monni:2011gb}, both in $\ee$ annihilation processes in QCD. Within the framework of Soft and Collinear Effective Theory (SCET)~\cite{Bauer:2000yr,Bauer:2001yt}, the N$^{3}$LL resummation for various event/jet variables have been performed~\cite{Chien:2010kc,Abbate:2010vw} and used, after matching to NNLO, for a precise determination of the coupling constant $\as$. The extracted value is consistent with the world average with significant improvements in the scale uncertainty.

At hadron colliders, what one often measures instead is jets, which only occupy
patches of the phase space. The corresponding jet shape variables are thus
non--inclusive, or non--global, and the resummation becomes highly non--trivial
even at NLL level. Consider, for example, measuring the normalised invariant
mass, $\rho$, of a subset of high--$\pt$ jets in multijet events. A veto is
applied on final state soft activity to keep the jet multiplicity fixed.
Jets are only defined through a jet algorithm, which generally depends on
some parameters such as the jet size $R$~\cite{Cacciari:2005hq}. We are thus
faced with a multi--scale ($\rho$, hard scale, veto, jet size) problem where
potentially large logs in the ratios of these scales appear. In addition to the
Sudakov leading logs, $\as^{n} \ln^{n+1}\rho$, coming from independent primary
gluon emissions, there are large subleading non--global logs (NGLs) of the form
$\as^{n} \ln^{n} (a/b)$, where $a$ and $b$ are two different scales, coming from
secondary~\footnote{These are emissions that are not radiated off primary hard
legs.} correlated gluon emissions.

We argued in~\cite{Banfi:2010pa} that in the narrow well--separated jets limit,
the non--global structure of the $\rho$ distribution, at hadronic colliders,
becomes much like that of $\ee$ hemisphere jet mass~\cite{Dasgupta:2001sh}. This
is mainly due to the fact that non--global logs arise predominantly near the
boundaries of individual jets. We had therefore considered $\ee$ dijet events
where only one of the jets is measured while the other is left unmeasured. We
found, in the anti--$\kt$ algorithm~\cite{Cacciari:2008gp}, NGLs in the ratio
$\rho Q/2 R^{2} \Eo$ as well as $2\Eo/Q$ where $\Eo$ and $Q$ are the veto and
hard scale respectively. These logs were completely missed out
in~\cite{Ellis:2009wj,Ellis:2010rwa}. The resummation of these NGLs to
all--orders had been approximated to that of the hemisphere
mass~\cite{Dasgupta:2001sh} up to terms vanishing as powers of $R$. Furthermore,
we pointed out, by explicitly computing the jet mass (without jet veto) distribution under clustering, that different jet definitions differ at NLL due to clustering--induced large logs. Here we compute these logs, which we refer to as \emph{clustering logs} (CLs), for the jet mass with a jet veto distribution.

Within the same context of $\ee$ multijet events, Kelley \emph{et
al.}~\cite{Kelley:2011tj} (version $1$) proposed that if one measures the masses of the two
highest--energy jets, instead of a single highest--energy jet as done
in~\cite{Banfi:2010pa}, then the resulting distribution is free from
NGLs. This is clearly not correct since the latter shape
observable, which we shall refer to, following~\cite{Kelley:2010qs}, as
\emph{threshold thrust}~\footnote{This name is more appropriate at hadron
colliders where at threshold the final state jets are back--to--back and there
is no beam remnant~\cite{Kelley:2010qs}.}, is still non--global. To clearly see
this consider, for example, the following gluonic configuration in $\ee$ dijet
events at $\Or(\as^{2})$. A gluon $k_{1}$ is emitted by hard eikonal legs into
the interjet energy region, $\Omega$. $k_{1}$ then emits a softer gluon $k_{2}$
into, say the quark jet region. This configuration then contributes to the quark
jet mass. The corresponding virtual correction, whereby gluon $k_{2}$ is
virtual, does not, however, contribute to the quark jet mass. Hence, upon adding
the two contributions one is left with a real--virtual mis--cancellation
resulting in logarithmic enhancement of the jet mass distribution. The latter is
what we refer to as  NGLs. The other, antiquark, jet receives identical enhancement. Thus
the sum of the invariant masses of the two jets does indeed contain NGLs
contribution. The latter is actually twice that of the single jet mass found
in~\cite{Banfi:2010pa}. 

Moreover, the authors of~\cite{Kelley:2011tj} (version $1$) claimed
that the anti--$\kt$~\cite{Cacciari:2008gp} and Cambridge--Aachen
(C--A)~\cite{Ellis:1993tq} jet algorithms only differ at NNLL for the threshold
thrust~\footnote{This claim has been removed from version $2$.}. From our
calculation in~\cite{Banfi:2010pa} for the jet mass, which
is not -with respect to clustering- much different from the threshold thrust, we know that the latter statement is incorrect. Nonetheless, an explicit proof will be presented below. Now, what is interesting in~\cite{Kelley:2011tj} and triggers the current work, is that the total differential threshold thrust distribution computed in the C--A algorithm and which contains neither NGLs nor CLs contributions, seemed to somehow agree well with next--to--leading (NLO) program \texttt{EVENT2}~\cite{Catani:1996jh}.

In this paper we shall shed some light on the result of~\cite{Kelley:2011tj} by
considering the individual colour, $\CF^{2}, \CF\CA$ and $\CF\TF\nf$, contributions to the total differential distribution as well as the effect of C--A clustering. We
show that at $\O(\as^{2})$ both NGLs and CLs are present and that
the above agreement with \texttt{EVENT2} is, on one side merely accidental~\footnote{As we shall see in sec.~\ref{sec.numerical_results}, while individual colour contributions do not agree with \texttt{EVENT2} their sum does, but only in the shape variable range and for the jet--radius considered in~\cite{Kelley:2011tj}. Outside the latter range or for other smaller jet--radii they do not agree.}, and on the other side due to the fact that the interval of the threshold thrust considered in~\cite{Kelley:2011tj} does not correspond to the asymptotic region where large logs are expected to dominate. The current work may be regarded as an extension to~\cite{Banfi:2010pa}. It includes: (a) computing the full $R$ dependence of the leading NGLs coefficient in the anti--$\kt$, (b) computing the small $R$ approximation of the latter as well as the leading CLs coefficient in the C--A algorithm and (c) checking our findings, as well as those of~\cite{Banfi:2010pa}, against \texttt{EVENT2}. It turns out, from the latter comparison, that the above approximation is actually valid for quite large values of $R$.

While the current paper was in preparation, a paper by Hornig \emph{et al.}~\cite{Hornig:2011tg} appeared in arXiv which studied NGLs in various jet algorithms, including anti--$\kt$ and C--A, within SCET. On the same day, Kelley \emph{et al.} published version $2$ of~\cite{Kelley:2011tj} in which they realised that this distribution is not actually free of NGLs and computed the corresponding coefficient in the anti--$\kt$ algorithm. Our findings on NGLs, which were independently derived using a different approach to both papers, confirm the results of both SCET groups. Clustering effects on primary emission sector are unique to this paper.

The organisation of this paper is as follows. In sec.~\ref{sec.fixed_order_1} we
compute the full logarithmic part of the LO threshold thrust distribution. We
then consider, in sec.~\ref{sec.fixed_order_2}, the fixed--order NLO 
distribution in the eikonal limit and compute the NGLs coefficient, in
both anti--$\kt$ and C--A jet algorithms. In the same section we derive an
expression for the CLs' first term as well. Note that our calculations for the C--A algorithm are performed in the small $R$ limit. Sec.~\ref{sec.resummation_in_QCD} is devoted to LL resummation of our jet shape including an exponentiation of the NGLs' and CLs' fixed--order terms. The latter exponentiation suffices for our purpose in this paper, which is to compare the analytical distribution with \texttt{EVENT2} at NLO. It
also provides a rough estimate of the size and impact of NGLs and CLs on the
total resummed distribution. In appendix~\ref{app.tw_in_SCET}, the corresponding
resummation in SCET~\cite{Kelley:2011tj,Kelley:2010qs,Becher:2008cf} is
presented. Numerical distributions of the threshold thrust obtained using the
program \texttt{EVENT2} are compared against analytical results and the findings discussed in sec.~\ref{sec.numerical_results}. In light of this discussion, we draw our main conclusions in sec.~\ref{sec.conclusion}.

\section{Fixed--order calculations: $\Or(\as)$} \label{sec.fixed_order_1}

After  briefly reviewing the definition of the threshold thrust observable, or
simply the jet mass with a jet veto, presented
in~\cite{Kelley:2011tj,Kelley:2010qs}, a general formula for sequential
recombination jet algorithms is presented. We then move on to compute the LO
integrated  distribution of this shape variable. At this order, all jet
algorithms are identical. Note that partons (quarks and gluons) are assumed
on--mass shell throughout.

\subsection{Observable and jet algorithms definitions} \label{subsec.observable_def}

Consider $\ee$ annihilation into multijet events. First, cluster events into
jets of size (radius) $R$ with a jet algorithm. After clustering, label the
momenta of the two hardest jets $p_{R}$ and $p_{L}$ and the energy of the third
hardest jet $E_{3}$. The threshold thrust is then given by the sum of the two
leading jets' masses after events with $E_{3} > \Eo$ are vetoed~\cite{Kelley:2011tj},
\begin{equation}\label{eq.tau_omega}
 \tauo = \frac{m^{2}_{R} + m_{L}^{2}}{Q^{2}} = \frac{\rho_{R} + \rho_{L}}{4}.
\end{equation}
$\rho_{R}$ and $\rho_{L}$ are the jet mass fractions for the two leading jets
respectively. We have shown in~\cite{Banfi:2010pa} that the single jet mass
fraction, $\rho$, is a non--global shape variable. Thus $\tauo$ must obviously
be a non--global variable too. 

A general form of sequential recombination algorithms at hadron colliders is
presented in~\cite{Cacciari:2005hq}. The adopted version for $\ee$ machines may
be summarised as follows~\cite{Cacciari:2005hq}: Starting with a list of final
state pseudojets with momenta $p_{i}$~\footnote{$p_{i}^{\mu}$ may be the momenta
of individual particles or each $p_{i}^{\mu}$ may be the total momentum of the
particles whose paths are contained in a small cell of solid angle about the
interaction point, as recorded in individual towers of a hadron calorimeter.},
energies $E_{i}$ and angles $\theta_{i}$ w.r.t. c.m frame, define the distances
\begin{equation}\label{jet_alg_def_theta}
d_{ij} = \min\left(E_{i}^{2 p},E_{j}^{2 p}\right)
\frac{2\left(1-\cos\theta_{ij}\right)}{R^{2}},\;\;\; d_{iB} = E_{i}^{2p},
\end{equation}
where $p$ can be any (positive or negative) continuous number. At a given stage
of clustering, if the smallest distance is $d_{ij}$ then $i$ and $j$ are
recombined together. Otherwise if the smallest distance is $d_{iB}$ then $i$ is
declared as a jet and removed from the list of pseudojets. Repeat until no
pseudojets are left. The recombination scheme we adopt here is the $E$--scheme,
in which pairs $(ij)$ are recombined by adding up their $4$--momenta. Two
pseudojets, $i$ and $j$, are merged together if
\begin{equation}\label{clust_cond}
  2(1-\cos\theta_{ij}) < R^{2}.
\end{equation}
The anti--$\kt$, C--A and $\kt$ algorithms correspond, respectively, to $p = -1,
p= 0$ and $p=1$ in eq.~\eqref{jet_alg_def_theta}. We shall only consider the
first two algorithms, anti--$\kt$ and C--A in this paper. Calculations for the
inclusive $\kt$ are identical to those for the C--A algorithm as shown
in~\cite{Banfi:2010pa}. With regard to notation, the jet--radius
in~\cite{Kelley:2011tj}, which we shall denote $\Rs$, is given in terms of $R$
by 
\begin{equation}\label{R_Rs_rel}
\Rs = R^{2}/4.
\end{equation}
Here we work with $\Rs$ instead of $R$.

To verify that the definition~\eqref{eq.tau_omega} is just the thrust in the
threshold (dijet) limit, hence the name, we begin with the general formula of
the thrust,
\begin{equation}\label{thrust_def}
 \tau = 1- \max_{\uv} \frac{\sum_{i} |\mathbf{p}_{i}.\uv |}{\sum_{i}
|\mathbf{p}_{i}|},
\end{equation}
where the sum is over all final state $3$--momenta $\mathbf{p}$ and the maximum
is over directions (unit vectors) $\uv$. In the threshold limit, enforced
by applying a veto on soft activity, $\ee$ annihilates into two back--to--back
jets and the \emph{thrust axis}, the maximum $\uv$, coincides with jet
directions. At LO, an emission of a single gluon, $k$, that is collinear to, and
hence clustered with say, $p_{R}$, produces the following contribution to the
thrust
\begin{equation}\label{eq.T^2}
 \tau \simeq \frac{E_{R} \omega}{Q}(1-\cos\theta_{k p_{R}}) + \frac{E_{L}
\omega}{Q}(1-\cos\theta_{k p_{L}}) + \frac{\omega^{2}}{Q^{2}}(1-\cos\theta_{k
p_{R}})(1-\cos\theta_{k p_{L}}),
\end{equation}
where $E_{R(L)}$ is the energy of the hard leg $p_{R(L)}$, $\omega$ the gluon's
energy and we have discarded an $\Or(\tau^{2})$ term. Recalling that the first two
terms in the RHS of eq.~\eqref{eq.T^2} are just the mass fractions $\rho_{R}$
and $\rho_{L}$, respectively, at LO and neglecting the third term (quadratic in
$\omega$) one concludes that 
\begin{equation}\label{tau_tauo}
\tau \simeq \tauo. 
\end{equation}
This relation can straightforwardly be shown to hold to all--orders.

\subsection{LO distribution}\label{sec.Fixed-PT-antikt}

In~\cite{Banfi:2010pa} we computed the LO distribution of the jet mass fraction,
$\rho$, in the small $R$ ($\Rs$) limit using the matrix--element squared in the
eikonal approximation. In this section, we use the full QCD matrix--element to
restore the complete $\Rs$ dependence of the singular part of the $\tauo$
distribution. The general expression for the integrated and normalised $\tauo$
distribution, or equivalently the $\tauo$ shape fraction, is given by
\begin{equation}\label{S1_tauo_dist}
\Sigma(\tauo,\Eo) = \int_{0}^{\tauo} \d\tauo' \int_{0}^{\Eo} \d
E_{3}\;\frac{1}{\sigma} \frac{\d^{2} \sigma}{\d\tauo' \d E_{3}},
\end{equation}
where $\sigma$ is the total $\ee \rightarrow $ hadrons cross--section. The
perturbative expansion of the shape fraction $\Sigma$ in terms of QCD coupling $\as$ may
be cast in the form
\begin{equation}\label{Sigma_PT}
\Sigma = \Sigma^{(0)} + \Sigma^{(1)} + \Sigma^{(2)} + \cdots,
\end{equation}
where $\Sigma^{(0)}$ refers to the Born contribution and is equal to $1$. The
derivation of the first order correction, $\Sigma^{(1)}$, to the Born
approximation is presented in appendix~\ref{app.LO_distr}. The final result
reads
\begin{multline}\label{R1_full-b}
\Sigma^{(1)}(\tauo,\Eo) = \frac{\CF \as}{2\pi} \left[-2\,\ln^{2}\tauo +\left(-3
+ 4\,\ln\frac{\Rs}{1-\Rs}\right)\,\ln\tauo \right] \Theta\left(\frac{\Rs}{1+\Rs}
-\tauo\right)+ \\+ \frac{\CF \as}{2\pi} \Bigg[ - 1 + \frac{\pi^{2}}{3} -
4\,\ln\frac{\Rs}{1-\Rs} \,\ln\frac{2\Eo}{Q} + f_{\Eo}(\Rs)\Bigg],
\end{multline}
where we have used eq.~\eqref{sig_had-sig_0} to change the normalisation in
eq.~\eqref{S1_tauo_dist} from $\sigma$ to $\sigma_{0}$. The reason for this
change is that the matrix--element we have used in \texttt{EVENT2} is normalised
to the Born cross--section~\footnote{Note that there are three sets of
matrix--elements included in the program, of which only one is not normalised to
the Born cross--section.}. The only difference between the two normalisations at
$\Or(\as)$ is in the one--loop constant. If we normalised to $\sigma$ we would
have found $\CF(-5/2+\pi^{2}/3)$ instead of $\CF(-1+\pi^{2}/3)$. The function
$f_{\Eo}(\Rs)$ is given by 
\begin{equation}\label{f_omeg}
f_{\Eo}(\Rs) = -2\,\ln\Rs\,\ln\frac{\Rs}{1-\Rs} + 2\,\Li_{2}(\Rs) -
2\,\Li_{2}(1-\Rs) + \frac{8 \Eo}{Q}\,\ln\frac{\Rs}{1-\Rs} +
\Or\left(\frac{\Eo^{2}}{Q^{2}}\right).
\end{equation}
Notice that eqs.~\eqref{R1_full-b} and~\eqref{f_omeg} are identical to eqs.~$(1)$ and~$(2)$ of~\cite{Kelley:2011tj} v$1$ and the sum of the $\as$ parts of eqs.~$(65)$ and $(66)$ in~\cite{Hornig:2011tg} provided that the jet radius in the latter, which we refer to as $\bar{R}$, is related to $\Rs$ by: $\tan^{2}(\bar{R}/2) = \Rs/(1-\Rs)$. It is worthwhile to note that in the limit $\Rs \rightarrow 1/2$ the $\tauo$ distribution~\eqref{R1_full-b} reduces to the well known thrust distribution~\cite{GehrmannDeRidder:2007bj} with upper limit $\tau < 1/3$. For $\Rs < 1/2$ the threshold thrust distribution includes, in addition to thrust
distribution, the interjet energy flow distribution~\cite{Oderda:1998en} too,
\begin{equation}
\Sigma^{(1)}_{\mathrm{E\,flow}}(\Eo) = \frac{\CF \as}{2\pi} \left[-
4\,\ln\frac{\Rs}{1-\Rs} \ln\left(\frac{2\Eo}{Q}\right) +
\Or\left(\frac{\Eo}{Q}\right) \right],
\end{equation}
Here the interjet region (rapidity gap), referred to in literature as $\deta$,
is defined by the edges of the jets. Specifically, it is related to the
jet--radius $\Rs$ by
\begin{equation}\label{eta-Rs}
\deta = -\ln\left(\frac{\Rs}{1-\Rs}\right).
\end{equation}

The important features of the $\tauo$ distribution that are of concern to the
present paper are actually contained in the second order correction term
$\Sigma^{(2)}$, which we address in the next section.
 
\section{Fixed--order calculations: $\Or(\as^{2})$}\label{sec.fixed_order_2}
 
We begin this section by recalling the formula of the matrix--element squared
for the $e^{+}e^{-}$ annihilation into two gluons, $\ee \longrightarrow q(p_{a})
+ \qbar(p_{b}) + g_{1}(k_{1})+ g_{2}(k_{2})$ in the eikonal approximation. Let
us first define the final state partons' $4$--momenta as
\begin{eqnarray}\label{4-momenta}
\nonumber p_{a} &=& \frac{Q}{2}(1,0,0,1) ,\\
\nonumber p_{b} &=& \frac{Q}{2}(1,0,0,-1) ,\\
\nonumber k_{1} &=&  \omega_{1}(1,\sin\theta_{1}\cos\phi_{1},
\sin\theta_{1}\sin\phi_{1},\cos\theta_{1}) ,\\
k_{2} &=&  \omega_{2}(1,\sin\theta_{2}\cos\phi_{2},
\sin\theta_{2}\sin\phi_{2},\cos\theta_{2}).
\end{eqnarray}
where the angles $\theta_{i}$ are w.r.t. $p_{a}$ direction (which lies along the
z--axis) and we assume the energies to be strongly ordered: $Q \gg \omega_{1} \gg
\omega_{2}$. This is so that one can straightforwardly extract the leading NGLs. Contributions from gluons with energies of the same order, $Q\gg \omega_{1} \sim \omega_{2}$, are subleading and hence beyond our control. The recoil effects are negligible in the former regime and are thus ignored throughout. The eikonal amplitude reads~\cite{QCD_collider},
\begin{equation}\label{W_2}
 S_{ab}(k_{1}, k_{2}) = \CF^{2} W_{P} + \CF\CA W_{S},
\end{equation}
where $W_{P}$ and $W_{S}$ stand for primary and secondary emission amplitudes
respectively. If we define the antenna function $w_{ij}(k) = 2(ij)/(ik)(kj)$
then the latter amplitudes are given by
\begin{equation}\label{W_P}
W_{P} = w_{ab}(k_{1}) w_{ab}(k_{2}) = \frac{16}{\omega_{1}^{2}\omega_{2}^{2}
\sin\theta_{1}^{2}\sin\theta_{2}^{2}},
\end{equation}
and
\begin{eqnarray}\label{W_S}
\nonumber W_{S} &=& \frac{w_{ab}(k_{1})}{2} \left[w_{a1}(k_{2})+w_{b1}(k_{2}) -
w_{ab}(k_{2})\right],
\\
&=& \frac{8}{\omega_{1}^{2}\omega_{2}^{2} \sin\theta_{1}^{2}
\sin\theta_{2}^{2}}\left[\frac{1 - \cos\theta_{1}
\cos\theta_{2}}{1-\cos\theta_{12}} - 1\right],\label{Ws_theta}
\end{eqnarray}
For completeness, the two--parton phase space is given by
\begin{equation}\label{PS_2}
\d\Phi_{2}(k_{1}, k_{2}) = \left[\prod_{i=1}^{2} \omega_{i}\d\omega_{i}
\frac{\sin\theta_{i}\d\theta_{i} \d\phi_{i}}{2\pi} \right]
\left(\frac{\as}{2\pi}\right)^{2},
\end{equation}
It is worth noting that the primary emission, $W_{P}$,
contribution to the $\tauo$ distribution is only fully accounted for by the
single--gluon exponentiation in the anti--$\kt$ algorithm case. If the final
state is clustered with a jet algorithm other than the latter, $W_{p}$
integration over the modified phase space, due to clustering, leads to (see
below) new logarithmic terms that escape the naive single--gluon exponentiation.
On the other hand, the secondary amplitude $W_{S}$ contribution is completely
missing from the latter Sudakov exponentiation in both algorithms.

First we outline the full $\as^{2}$ structure of the $\tauo$ distribution up
to NLL level in the anti--$\kt$ including the computation of the NGLs
coefficient. After that, we investigate the effects of final state partons'
clustering on both primary and secondary emissions. The C--A
algorithm is taken as a case study to illustrate the main points. Calculations
where the final state is clustered with other jet algorithms should proceed in
an analogous way to the C--A case.

\subsection{$\tauo$ distribution in the anti--$\kt$ algorithm}\label{subsec.FO_anti-kt}

The anti--$\kt$ jet algorithm works, in the soft limit, like a perfect cone.
That is, a soft gluon $k_{i}$ is clustered to a hard parton $p_{j}$ if it is
within an angular distance $2\sqrt{\Rs}\, ( = R)$, from the axis defined
by the momentum of the latter. This feature of the algorithm greatly simplifies
both fixed--order and resummation calculations. Considering all possible angular
distances between $(k_{1},k_{2})$ and $(p_{a},p_{b})$ we compute below the
corresponding contributions to primary and secondary pieces of the $\tauo$
distribution. Note that we use LL and NLL to refer to leading and
next--to--leading logs of $\tauo$ (and not $2\Eo/Q$) in the exponent of the
resummed distribution (discussed in sec.~\ref{sec.resummation_in_QCD}).

\subsubsection{$\CF^{2}$ term} \label{subsec.anti-kt_CF2}

The LL contribution to the $\tauo$ distribution comes from diagrams
corresponding to two--jet final states. That is diagrams where both real gluons,
$k_{1}$ and $k_{2}$, are clustered with the hard partons $p_{a}$ and $p_{b}$.
Diagrams where one of the two gluons is in the interjet region, and hence not
clustered with either hard parton, contribute at NLL level. Other gluonic
configurations lead to contributions that are beyond our NLL control and thus
not considered. The $\CF^{2}$ part of the $\Or(\as^{2})$ $\tauo$
distribution may be found by expanding the exponential of the LO
result~\eqref{R1_full-b}. The full expression including the
running coupling at two--loop in the $\overline{\mathrm{MS}}$ will be presented in sec.~\ref{sec.resummation_in_QCD}. For the sake of comparison to the clustering case, we only report here the the LL term, which reads
\begin{equation}\label{FO_2_CF2}
\Sigma^{(2)}_{P}(\tauo,\Eo) = 2\,\CF^{2}
\left(\frac{\as}{2\pi}\right)^{2}\,\ln^{4}(\tauo).
\end{equation} 

Next we consider the derivation of the $\CF\CA$ contribution to the  $\tauo$
distribution including the full jet--radius dependence.

\subsubsection{$\CF\CA$ term and NGLs}\label{subsec.anti-kt-NGL}

In the anti--$\kt$ algorithm the non--global logarithmic contribution to the
$\tauo$ distribution is simply the sum of that of the single jet mass fraction,
$\rho$, with a jet veto distribution studied in~\cite{Banfi:2010pa}~\footnote{Here we go beyond the small $\Rs$ approximation assumed in~\cite{Banfi:2010pa}.}. This is in
line with the near--edge nature of non--global enhancements. In two--jet events,
the well separated~\footnote{such that the jet--radius is much smaller than the
jets' separation; $\Rs \ll (1- \cos\theta_{ij})$, where $\theta_{ij}$ is the
angle between jets $i$ and $j$.} jets receive the latter enhancements
independently of each other.
\begin{figure}
\centering
\includegraphics[width=15cm]{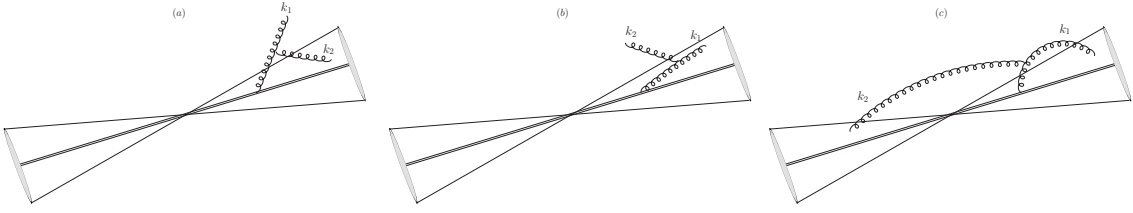}
\caption{Schematic representation of gluonic arrangement giving rise to NGLs. We
have only shown the NGLs contributions to the $p_{R}$--jet. Identical
contributions apply to the $p_{L}$--jet.}
\label{fig.NGLs_akt}
\end{figure}
Possible final state gluonic arrangements relevant to NGLs at second order are
depicted in fig.~\ref{fig.NGLs_akt}. The all--orders resummed NGLs distribution
may be written in the form~\cite{Dasgupta:2001sh}
\begin{equation}\label{S_t_gen}
S(t) = 1+ S_{2}\,t^{2} + \cdots = 1+ \sum_{n=2}^{} S_{n}\,t^{n},
\end{equation}
with $t$ being the evolution parameter defined in terms of the coupling $\as$ by
\begin{eqnarray}\label{t_param_akt}
 \nonumber t &=& \frac{1}{2\pi} \int_{\kt^{\min}}^{\kt^{\max}} \frac{\d\kt}{\kt}
\;\as(\kt),\\
 &=& \frac{\as}{2\pi}\; \,\ln\left(\frac{\kt^{\max}}{\kt^{\min}}\right),
\end{eqnarray}
where the exact form of the upper and lower limits, $\kt^{\max}$ and
$\kt^{\min}$, depend on the gluonic configuration and the second line
in~\eqref{t_param_akt} assumes a fixed coupling. To make contact with interjet
energy flow calculations~\cite{Dasgupta:2002bw,Appleby:2002ke}, we work in this
particular section with hadronic variables $(k_{t},\eta, \phi)$ instead of $\ee$
variables $(E,\theta, \phi)$. The pseudo--rapidity $\eta$ and transverse
momentum $\kt$ (both measured w.r.t. incoming beam direction) are related,
respectively, to the angle and energy by~\footnote{Otherwise, one can redefine
the partons' $4$--momenta in terms of $\eta$ and $\kt$ and use the antenna
function expressions of $W_{P}$ and $W_{S}$ to rewrite them in terms of the hadronic variables.}
\begin{equation}
\eta = - \ln\left(\tan\frac{\theta}{2}\right),\;\;\; E =\kt \cosh(\eta)
\end{equation}
Using the secondary emissions eikonal amplitude~\eqref{W_S} in terms of the new
variables, the NGLs coefficient $S_{2}$ reads
\begin{equation}\label{S2_akt_gen}
S_{2} = -4 \CF\CA
\;\int\d\Phi^{(2)}\;\left[\frac{\cosh(\eta_{1}-\eta_{2})}{\cosh(\eta_{1}-\eta_{2
}) - \cos(\phi_{1} -\phi_{2})} -1\right],
\end{equation}
where the phase space measure, $\d\Phi^{(2)}$, is of the general form given in
eq.~\eqref{PS_2} with the $\kt$ integrals included in the definition of
$t$~\eqref{t_param_akt} and new restrictions coming from the jet shape
definition. For configuration $(a)$ in fig.~\ref{fig.NGLs_akt}, it reads
\begin{equation}\label{S2_PS_a}
\d\Phi^{(2)}_{a} = \int_{-\frac{\deta}{2}}^{\frac{\deta}{2}} \d\eta_{1}
\frac{\d\phi_{1}}{2\pi} \times 2 \int_{\frac{\deta}{2}}^{+\infty}\d\eta_{2}
\frac{\d\phi_{2}}{2\pi}\; \Theta\left( \ln\frac{\ktt}{Q\tauo} - \eta_{2}\right)
\Theta\left(\Eo - \kto\cosh(\eta_{1})\right), 
\end{equation}
where the interjet (gap) region, $\deta$ is given in eq.~\eqref{eta-Rs}. Due to
boost invariance of rapidity variables the latter region has been centred at
$\eta=0$. Moreover, the factor $2$ in~\eqref{S2_PS_a} accounts for the
$p_{L}$--jet contribution. Since neither the integrand nor the integral measure in
eq.~\eqref{S2_akt_gen} depends explicitly on the azimuthal angles ($\phi_{1}$, $\phi_{2}$), we use our freedom to set $\phi_{1} = 0$, average over $\phi_{2}$ and then
perform the rapidity integration. The resultant expression for $S_{2}$ in configuration $(a)$ at the limit $\tauo \rightarrow 0$ reads,
\begin{equation}\label{S2_akt_a1}
 S_{2,a} = -4\CF\CA \left[\frac{\pi^{2}}{12} +
\deta^{2} - \deta\,\ln\left(e^{2\deta}-1\right) -\frac{1}{2}
\Li_{2}\left(e^{-2\deta}\right) -\frac{1}{2} \Li_{2}\left(1-e^{2\deta}\right)
\right]
\end{equation}
An identical expression was found for the NGLs' coefficient in the interjet
energy flow distribution~\cite{Dasgupta:2002bw}~\footnote{Our jet--radius,
$\Rs$, is given in terms the parameter $c$, used in~\cite{Dasgupta:2002bw}, by
the relation: $1 - c = 2\Rs$.}. The fact that $S_{2,a}$ is the same for $\tauo$
and interjet energy flow distributions means that the NGLs' coefficient only
depends on the geometry of the phase space and not on the observable itself.
This is of course only true in the limit where the jet shape variable goes to
zero. The difference between the jet shape variables amounts only to a
difference in the logarithm's argument. 

It should be understood that there are $\Theta$--function constraints
on $\kto$ and $\ktt$ resulting from rapidity integrations not explicitly shown
in eq.~\eqref{S2_akt_a1}. Performing the remaining trivial $\kt$
integrals yields
\begin{equation}\label{t_param_a}
 t_{a}^{2} = \left(\frac{\as}{2\pi}\right)^{2}\;\ln^{2}\left(\frac{2\Eo\,\Rs}{Q\tauo}\right)  \;\Theta\left(\frac{2\Eo}{Q} - \frac{\tauo}{\Rs}\right),
\end{equation}
where a factor of $1/2$ has been absorbed in $S_{2,a}$~\eqref{S2_akt_a1}. 

Now consider configuration $(b)$ in fig.~\ref{fig.NGLs_akt}. Adding up the
corresponding virtual correction, one obtains the following phase space
constraint
\begin{equation}\label{PS_const_b}
\Theta\left(\eta_{1} -\ln\left(\frac{\kto}{Q\tauo} \right) \right) \Theta\left(\ktt -
\frac{\Eo}{\cosh(\eta_{2})}\right).
\end{equation}
The phase space measure $\d\Phi_{b}^{(2)}$ is analogous to $\d\Phi_{a}^{(2)}$
in~\eqref{S2_PS_a} with $1 \leftrightarrow 2$ and the two $\Theta$--functions
in~\eqref{S2_PS_a} replaced by those in eq.~\eqref{PS_const_b}. The limits on
$\eta_{1}$ are then $+\infty > \eta_{1} > \max[\deta/2,\ln(\kto/Q\tauo)]$. If we
impose the constraint given in eq.~\eqref{t_param_a}, i.e, $2\Eo/Q \gg
\tauo/\Rs$, then the lower limit becomes $\eta_{1} > \ln(\kto/Q\tauo)$. The NGLs
coefficient $S_{2,b}$ thus reads
\begin{equation}\label{S2_akt_b}
S_{2,b} = -4\CF\CA \int_{\ln\frac{\kto}{Q\tauo}}^{+\infty} \d\eta_{1}
\int_{-\frac{\deta}{2}}^{\frac{\deta}{2}}\d\eta_{2}
\left[\coth(\eta_{1}-\eta_{2}) -1\right],
\end{equation}
where we have averaged the eikonal amplitude $W_{s}$ over $\phi_{2}$ and moved
$\kt$s' $\Theta$--functions onto the integral of the evolution parameter $t_{b}$, which
is given at $\as^{2}$ by
\begin{equation}\label{t_param_akt_b}
t_{b}^{2} = \left(\frac{\as}{2\pi}\right)^{2} \ln^{2}\left(\frac{2\Eo}{Q}\right)
\Theta\left(\frac{2\Eo}{Q} - \frac{\tauo}{\Rs}\right).
\end{equation}
The $S_{2, b}\, t_{b}^{2}$ contribution is then beyond our NLL accuracy. In
fact, $S_{2,b}$ vanishes in the limit $\tauo \rightarrow 0$ as can be seen from
eq.~\eqref{S2_akt_b}.

The last contribution to NGLs at $\Or(\as^{2})$ comes from configuration $(c)$ in
fig.~\ref{fig.NGLs_akt}. Upon the addition of the virtual correction, one is
left with the constraint
\begin{equation}\label{PS_const_c}
\Theta\left(Q\tauo - \kto e^{-\eta_{1}} \right) \Theta\left(\ktt e^{+\eta_{2}} -
Q\tauo\right).
\end{equation}
The corresponding NGLs coefficient and evolution parameter read
\begin{eqnarray}\label{S2_akt_c}
S_{2,c} &=& -4\CF\CA \int_{\max\left[\ln\frac{\kto}{Q\tauo}, \frac{\deta}{2}\right]}^{+\infty}\d\eta_{1}
\int_{-\ln\frac{\ktt}{Q\tauo}}^{-\frac{\deta}{2}} \d\eta_{2}
\left[\coth(\eta_{1} -\eta_{2}) -1\right],\label{S2_b_limit}
\\
t_{c}^{2} &=& \left(\frac{\as}{2\pi}\right)^{2}
\ln^{2}\left(\tauo\,e^{\deta/2}\right),\label{t_param_akt_c}
\end{eqnarray}
Since we have assumed strong ordering, $\kto \gg \ktt$, then the lower limit of $\eta_{1}$ in~\eqref{S2_b_limit} is $\ln(\kto/Q\tauo)$. Consequently the coefficient $S_{2,c}$ vanishes in the limit $\tauo \rightarrow0$. For this reason, this configuration will not be considered. 

We conclude that in the regime $2\Eo/Q \gg \tauo/\Rs$, the only non--vanishing
contribution to the NGLs comes from the phase space configuration $(a)$. Other
configurations, $(b)$ and $(c)$, vanish in the limit $\tauo \rightarrow 0$. Hence 
\begin{equation}\label{S2_akt_tot}
 S_{2} = S_{2,a},\;\;\; t = t_{a}.
\end{equation}
In fig.~\ref{fig.NG_coff_CA} we plot $S_{2}$ as a function of the jet--radius $\Rs$. At the asymptotic limit $\deta \rightarrow +\infty$ (or equivalently $\Rs \rightarrow 0$) $S_{2}$ saturates at $ -\CF\CA\; 2\pi^{2}/3$. This value (or rather half of it) is used as an approximation to $S_{2}$ in~\cite{Banfi:2010pa}. From eq.~\eqref{S2_akt_a1}, we can see that the correction to such an approximation is less than $10\%$ for jet--radii smaller than $\Rs \sim 0.28$, which is equivalent to $R \sim 1$.  Furthermore, Eq.~\eqref{S2_akt_a1} confirms the claim made in the same paper that NGLs do not get eliminated when the jet--radius approaches zero. One may naively expects that when the jet size shrinks down to $0$ ($\Rs\rightarrow 0$) there is no room for gluon $k_{2}$ to be emitted into. This means that $\tauo$ becomes inclusive and hence $S_{2}$ vanishes. To the contrary, $S_{2}$ reaches its maximum in this limit.

Few important points to note:
\begin{itemize}
  \item If we choose to order the energy scales in the $\Theta$--functions of~\eqref{t_param_a} and~\eqref{t_param_akt_b} the opposite way, i.e, $2\Eo/Q \ll \tauo/\Rs$ then configuration $(b)$ becomes leading, in NGLs, while the contribution from configuration $(a)$ vanishes. That is $t_{b}^{2}$ in eq.~\eqref{t_param_akt_b} becomes
   \begin{equation}\label{t_param_akt_b_no-ordering}
     t_{b}^{2} = \left(\frac{\as}{2\pi}\right)^{2}\ \ln^{2}\left(\frac{2\Eo\Rs}{Q\tauo}\right)\,\Theta\left(\frac{\tauo}{\Rs} - \frac{2\Eo}{Q}\right).
   \end{equation}
  and $S_{2,b} = S_{2,a}$ in eq.~\eqref{S2_akt_a1}. We do not consider this regime here though. 

  \item If, on the other hand, we do not restrict ourselves to any particular ordering of the scales, as it is done in Refs.~\cite{Hornig:2011tg} and~\cite{Kelley:2011tj}, then both configurations $(a)$ and $(b)$ would contribute to the leading NGLs. Adding up $t_{a}^{2}$, in~\eqref{t_param_a}, and $t_{b}^{2}$, in~\eqref{t_param_akt_b_no-ordering}, the $\Theta$--functions sum up to unity and one recovers the result reported in the above mentioned references. Notice that it is a straightforward exercise to show that eq.~\eqref{S2_akt_a1} is equal to $f_{\mathrm{OL}} + f_{\mathrm{OR}}$ given in eq.~$(28)$ of~\cite{Hornig:2011tg} in the case where $R_{L} = R_{R} = R$ ($=\bar{R}$ given in sec.~\ref{sec.fixed_order_1}). Moreover, the coefficient $f^{\CA}_{\mathrm{NGL}}$ given in eq.~$(\mathrm{B}2)$ of~\cite{Kelley:2011tj} v$2$ is related to $S_{2,a}$ by $ f^{\CA}_{\mathrm{NGL}} = -8 \times S_{2,a}$.
 
 \item Setting the cut--off scale $\Eo \sim \tauo Q$ in $t_{a}$, eq.~\eqref{t_param_a}, and $t_{b}$, eq.~\eqref{t_param_akt_b_no-ordering}, would diminish NGLs coming from both configurations $(a)$ and $(b)$ and the threshold thrust becomes essentially a global observable. This is unlike the observation made in the study of the single jet with a jet veto distribution~\cite{Banfi:2010pa} where the above choice of $\Eo$ kills the NGLs near the measured jet but introduces other equally significant NGLs near the unmeasured jet.
\end{itemize}

In the next subsection we recompute both $\CF^{2}$ and $\CF\CA$ contributions to
the $\tauo$ distribution under the C--A clustering condition. For the $\CF\CA$
term, we only focus on configuration $(a)$ and do not attempt to address the
subleading contributions coming from configurations $(b)$ and $(c)$.

\subsection{$\tauo$ distribution in the C--A algorithm}\label{sec.FO_CA}

The definition of the C--A algorithm is given in eq.~\eqref{jet_alg_def_theta}
with $p = 0$. Unlike the anti--$\kt$ algorithm, which successively merges soft
gluons with the nearest hard parton, the  C--A algorithm proceeds by
successively clustering soft gluons amongst themselves. Consequently, a soft
parton may in many occasions be dragged into (away from) a jet region and hence
contributing (not contributing) to the invariant mass of the latter. The jet
mass, and hence $\tauo$, distribution is then modified. It is these
modifications, due to soft--gluons self--clustering, that we shall address
below.

Any clustering--induced contribution to the $\tauo$ distribution will only arise
from phase space configurations where the two soft gluons, $k_{1}$ and $k_{2}$,
are initially (that is, before applying the clustering) in different regions of
phase space. Configurations where both gluons are within the same jet region,
gluon $k_{1}$ is in one of the two jet regions and gluon $k_{2}$ is in the other
or both gluons are within the interjet region are not altered by clustering and
calculations of the corresponding contributions will yield identical results to
the anti--$\kt$ algorithm. We can therefore write the $\tauo$ distribution in
the C--A algorithm, at $\Or(\as^{2})$, as
\begin{equation}\label{Sig2_CA_akt}
\Sigma^{(2)}_{\ca}(\tauo,\Eo) = \Sigma^{(2)}_{\akt}(\tauo,\Eo) + \delta
\Sigma^{(2)}(\tauo,\Eo).
\end{equation}
It is the last term in eq.~\eqref{Sig2_CA_akt} that we compute in the present
subsection. Starting at configurations with two gluons in two different regions,
the jet algorithm either:
\begin{enumerate}[(A)]
  \item recombines the two soft gluons into a single parent gluon if the
clustering condition~\eqref{clust_cond} is satisfied. The latter parent gluon
will either be in one of the two jet regions or out of both of them (and hence
in the interjet region).
  \item or leaves the two gluons unclustered, if the clustering condition is not
satisfied. This case is then identical to the anti--$\kt$ one but with a more
restricted phase space. This restriction comes from the fact that for the two
gluons to survive the clustering they need to be sufficiently far apart.
Quantitatively, their angular separation should satisfy the relation
\begin{equation}\label{CA_12_clust_cond}
  (1-\cos\theta_{12}) > 2\Rs.
\end{equation}  
\end{enumerate}
Below, we examine the contributions from configurations (A) and (B) to the
$\CF^{2}$ and $\CF\CA$ colour pieces of the $\tauo$ distributions. All
calculations are performed in the small $\Rs$ approximation using the $\ee$ variables $(\omega,\theta, \phi)$.

\subsubsection{$\CF^{2}$ term}\label{subsec.CA-LL}

\begin{figure}[t]
\centering
\includegraphics[width=15cm]{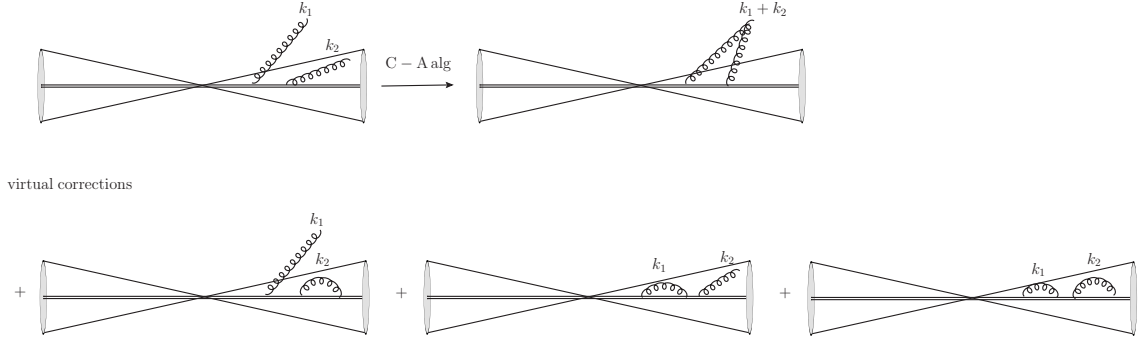}
\caption{A schematic representation of a three--jet final state after applying
the C--A algorithm on real emission along with virtual correction diagrams. The
two gluons are clustered in the E--Scheme (see sec.~\ref{sec.fixed_order_1}).
Identical diagrams hold for the $p_{L}$--jet.}
\label{fig.CLs_CF2}
\end{figure}
Consider the gluonic configuration in (A) where the harder gluon $k_{1}$ is in
the interjet region and the softer gluon $k_{2}$ is in the $p_{R}$--jet region. We account for the $p_{L}$--jet region through multiplying the final result
by a factor of two. Applying the C--A algorithm~\eqref{jet_alg_def_theta}, the
smallest distance is $d_{\min} = d_{12}$. Hence gluon $k_{1}$ pulls gluon
$k_{2}$ out of the $p_{R}$--jet region and form a third jet, as depicted in
fig.~\ref{fig.CLs_CF2}. The latter is then vetoed to have energy less than
$\Eo$. The corresponding clustering angular function, in the small angles limit,
reads
\begin{eqnarray}\label{clust_CA_CF2_theta}
\nonumber \Theta_{\ca}(1,2) &=& \Theta(\theta_{1}^{2}- 4\Rs) \Theta(4\Rs -
\theta_{2}^{2}) \Theta(\theta_{2}^{2} - \theta_{12}^{2}),
\\
\nonumber &=& \Theta(4 \theta_{2}^{2}\cos^{2}\phi_{2} -\theta_{1}^{2})
\Theta(\theta_{1}^{2}- 4\Rs) \theta(4\Rs - \theta_{2}^{2})
\Theta\left(\theta_{2}^{2} - \frac{\Rs}{\cos^{2}\phi_{2}}\right)
\Theta\left(\cos\phi_{2} - \frac{1}{2}\right).\\
\end{eqnarray}
Adding up the corresponding virtual corrections, where one or both of the gluons
are virtual, one obtains the following constraint on the phase space
\begin{equation}
\Theta(\Eo -\omega_{1} -\omega_{2}) -\Theta(\Eo -\omega_{1}) +
\Theta\left(\frac{\omega_{2}}{2Q} \theta_{2}^{2} - \tauo\right).
\end{equation}
Since we are working in the strong energy--ordered regime, $\omega_{1} \gg
\omega_{2}$, only the last $\Theta$--function survives. The new contribution to
the $\CF^{2}$ piece of the $\tauo$ distribution is then given by
\begin{eqnarray}\label{C2_CA}
\nonumber C_{2}^{P} t_{p}^{2} &=& 8 \int^{Q/2}_{\frac{Q\tauo}{2\Rs}}
\frac{\d\omega_{2}}{\omega_{2}}
\int^{Q/2}_{\omega_{2}}\frac{\d\omega_{1}}{\omega_{1}}
\int_{-\frac{\pi}{3}}^{\frac{\pi}{3}}\frac{\d\phi_{2}}{2\pi}
\int_{2\sqrt{\Rs}}^{2\theta_{2}\cos\phi_{2}} \frac{\d\theta_{1}}{\theta_{1}}
\int_{2\sqrt{\Rs}}^{\Rs/\cos\phi_{2}} \frac{\d\theta_{2}}{\theta_{2}},
\\
&=& 0.73 \;\CF^{2}\; \left(\frac{\as}{2\pi}\right)^{2}
\ln^{2}\left(\frac{\Rs}{\tauo}\right).
\end{eqnarray}
This result is identical to~\footnote{It is actually twice} that found in~\cite{Banfi:2010pa} for a single jet
mass (without a jet veto) distribution. The reason for this is that the
clustering requirement only affects the distribution to which the softest gluon
contributes. Which is in both cases the jet mass distribution.

\begin{figure}[t]
\centering
\includegraphics[width=15cm]{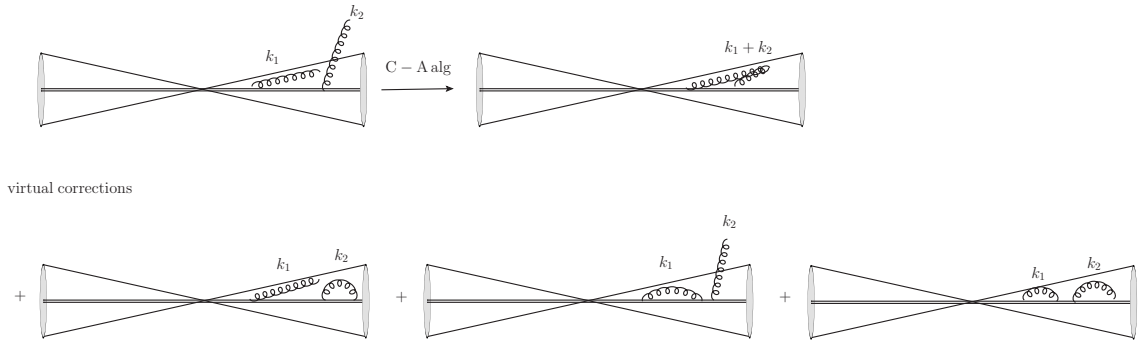}
\caption{A schematic representation of a two--jet final state after applying the
C--A algorithm on real emission along with virtual correction diagrams. The two
gluons are clustered in the E--Scheme (see sec.~\ref{sec.fixed_order_1}).
Identical diagrams hold for the left ($p_{L}$--) jet.}
\label{fig.CLs_2}
\end{figure}
The second possible configuration that corresponds to case (A) is where gluon
$k_{1}$ is in, say, the $p_{R}$--jet region and the softer gluon $k_{2}$ is in
the interjet region. If the two gluons are clustered, i.e, gluon $k_{1}$ pulls
in gluon $k_{2}$, then upon adding real emission and virtual correction
diagrams, depicted in fig.~\ref{fig.CLs_2}, one obtains the following phase
space constraint
\begin{equation}\label{PS_const_CA_CF2}
-\Theta\left(\tauo -\frac{\omega_{1}}{2Q} \theta_{1}^{2}\right)
\Theta\left(\frac{\omega_{2}}{2 Q} \theta_{2}^{2} - \tauo\right) +
\Theta(\omega_{2} - \Eo),
\end{equation}
where we have assumed small angles limit and employed the LL accurate
approximation
\begin{equation}
\Theta\left(\tauo - \frac{\omega_{1}}{2Q} \theta_{1}^{2} - \frac{\omega_{2}}{2Q}
\theta_{2}^{2}\right) \simeq \Theta\left(\tauo - \frac{\omega_{1}}{2Q}
\theta_{1}^{2} \right) \Theta\left(\tauo - \frac{\omega_{2}}{2Q}
\theta_{2}^{2}\right).
\end{equation}
Given the fact that $\omega_{1} \gg \omega_{2}$ and $\theta_{1}$ and
$\theta_{2}$ must be close to each other to be clustered, i.e, they should
satisfy condition~\eqref{CA_12_clust_cond}, then the first two
$\Theta$--functions in eq.~\eqref{PS_const_CA_CF2} are substantially suppressed
and one is only left with the veto on $\omega_{2}$. Applying the C--A algorithm
one obtains an identical clustering function to eq.~\eqref{clust_CA_CF2_theta}.
Hence the CLs' coefficient for this configuration is equal to $C_{2}^{P}$ given
in eq.~\eqref{C2_CA}. That is $C_{2}^{P} = 0.73\,\CF^{2}$. The evolution
parameter does however change. It is now given, at $\Or(\as^{2})$, by
\begin{equation}\label{C2_CA_b}
t_{p}^{' 2} = \left(\frac{\as}{2\pi}\right)^{2}
\ln^{2}\left(\frac{2\Eo}{Q}\right).
\end{equation}
This contribution is then beyond our NLL control. Note that the CLs contribution
in eq.~\eqref{C2_CA_b} is equal to what one would find for interjet energy flow
distribution provided that the rapidity gap is defined through eq.~\eqref{eta-Rs}.

Let us now turn to case (B) where the two gluons are not merged together. If
gluon $k_{1}$ is in the interjet region and gluon $k_{2}$ is in one of the two
jet regions then the corresponding phase space constraint reads
\begin{equation}\label{CF2_CA_B}
\Theta\left(\frac{2Q\tauo}{\omega_{2}} -\theta_{2}^{2}\right) \\
\Theta(\omega_{1} - \Eo) \left[1 - \Theta_{\ca}(1,2)\right].
\end{equation}
The limits on $\theta_{2}$--integral are then given by:
$\min(4\Rs,2Q\tauo/\omega_{2}) > \theta_{2}^{2}>0$. Imposing the constraint
$2\Eo/Q \gg \tauo/\Rs$, it is straightforward to see that the above constraint
yields NNLL contribution and thus beyond our control. Similarly, the
configuration where gluon $k_{1}$ is in the jet region and gluon $k_{2}$ is in
the interjet region yields subleading logs. 

Hence the $\CF^{2}$ piece of the clustering--induced correction term
$\delta\Sigma^{(2)}$, in eq.~\eqref{Sig2_CA_akt}, up to NLL, reads
\begin{equation}\label{Sig_CA_CF2}
\delta\Sigma^{(2)}(\tauo,\Eo) = C^{P}_{2}\,t_{p}^{2},
\end{equation}
Next we compute the $\CF\CA$ piece of $\delta\Sigma^{(2)}$.

\subsubsection{$\CF\CA$ term}\label{subsec.CA-NGL}

\begin{figure}[t]
	\centering
	\includegraphics[width=11cm]{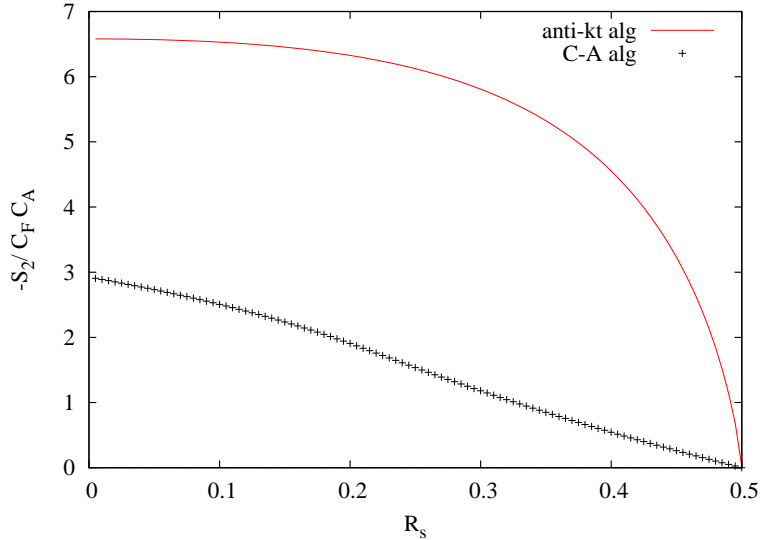}
	\caption{Non--global coefficient $S_{2}$ in the anti--$\kt$ and C--A algorithms.}
\label{fig.NG_coff_CA}
\end{figure}

Consider the gluonic configuration $(a)$ depicted in fig.~\ref{fig.NGLs_akt}.
Applying the C--A clustering algorithm on the latter yields two possibilities.
Namely the two gluons are either clustered or not. The former case completely
cancels against virtual corrections and thus does not contribute to NGLs. It is when the two gluons survive the
clustering, the latter case, that a real--virtual mismatch takes place and NGLs
are induced. The corresponding evolution parameter is equal to $t$ of the
anti--$\kt$ case, eq.~\eqref{S2_akt_tot}. The clustering condition is simply one minus that in
eq.~\eqref{clust_CA_CF2_theta}. The NGLs' coefficient can then be written, using
the eikonal amplitude~\eqref{Ws_theta}, as
\begin{equation}\label{S2_CA_theta}
S^{\ca}_{2} = S_{2} + \delta\Sigma^{(2)}_{\CF\CA}
\end{equation}
where $S_{2}$ is given in eq.~\eqref{S2_akt_tot} and
\begin{multline}
\delta\Sigma^{(2)}_{\CF\CA} = 8\,\CF\CA
\int_{\sqrt{\Rs}}^{2\theta_{2}\cos\phi_{2}}\frac{\d\theta_{1}}{\sin\theta_{1}}
\int_{\frac{\sqrt{Rs}}{\cos\phi_{2}}}^{2\sqrt{\Rs}}\frac{\d\theta_{2}}{
\sin\theta_{2}} \int_{-\pi/3}^{\pi/3}\frac{\d\phi_{2}}{2\pi}
\left[\frac{1-\cos\theta_{1}\cos\theta_{2}}{1-\cos\theta_{12}} -1\right] \times
\\ \times \Theta\left(\frac{\Rs}{\tauo\cos\phi_{2}} -
\frac{Q}{2\omega_{2}}\right),
\end{multline}
We can perform the $\theta_{1}$--integral analytically and then resort to
numerical methods to evaluate the remaining $\theta_{2}$ and $\phi_{2}$
integrals. The result, in terms of the jet--radius $\Rs$, is depicted in
Fig.~\ref{fig.NG_coff_CA}.
$- S^{\ca}_{2}$ saturates at around $0.44\times 2\pi^{2}/3\, \CF\CA\sim
2.92\, \CF\CA$, i.e, a reduction of about $55\%$ in $S_{2}$. This is due to
the fact that for the two gluons to survive clustering they need to be
sufficiently far apart ($\theta_{12} > R = 2\sqrt{\Rs}$). The dominant
contribution to $S_{2}$ comes, however, from the region of phase space where
the gluons are sufficiently close. This corresponds to the collinear region of
the matrix--element; $\theta_{1} \sim \theta_{2}$. Hence the further apart the
two gluons get from each other, the less (collinear) singular the matrix becomes
and thus the smaller the value of NGLs coefficient.

Note that the C--A coefficient $S_{2}^{\ca} = 2\times f^{\mathrm{C/A}}_{\mathrm{OR}}$, where $f^{\mathrm{C/A}}_{\mathrm{OR}}$ is given by eq.~$(38)$ in~\cite{Hornig:2011tg}, at least in the small jet--radius region. Noticeably, the two results coincide at both limits $\Rs \rightarrow 0$ and $\Rs \rightarrow 1/2$ (equivalently $R \rightarrow 0$ and $R \rightarrow \sqrt{2}$ in~\cite{Hornig:2011tg}). In fact, the coefficient $S_{2}^{\ca}$ is valid, as we shall see in sec.~\ref{sec.numerical_results}, for quite large jet--radii; up to $\Rs \sim 0.3$ (equivalent to $R \sim 1$ in~\cite{Hornig:2011tg}).

The fixed--order NLL logarithmic structure of the $\tauo$ distribution should by
now be clear for both jet algorithms. In order to assess the phenomenological impact of NGLs and clustering requirement on the final cross--section, it is necessary to perform
an all--orders treatment, which we do below.

\section{Resummation of $\tauo$ distribution}\label{sec.resummation_in_QCD}

Resummation, which is essentially the organisation of large logs arising from
soft and/or collinear radiation to all--orders, is based on the factorisation
property of the pQCD matrix--element squared for multiple gluon radiation. This
is only true for independent primary emissions though. Including secondary
correlated emissions, the picture dramatically changes and the resummation can
only be performed at some limits, eg. large--$\Nc$ limit~\cite{Banfi:2005gj}. In
the standard method~\cite{Collins:1984kg,Catani:1992ua,Bonciani:2003nt},
resummation is carried out in Mellin (Laplace) space instead of momentum space.
Only at the end does one transform the result back to the momentum space through
(inverse Mellin transform),
\begin{equation}\label{inv_Mell_tauo}
\Sigma_{P}(\tauo, \Eo) = \int \frac{\d\nu}{2\imath\pi\nu}\; e^{\nu\tauo}\; \int
\frac{\d\mu}{2\imath\pi\mu}\; e^{\mu\Eo}\;\widetilde{\Sigma}_{P}(\nu^{-1},
\mu^{-1}),
\end{equation}
where $P$ stands for primary emission. With regard to non--global observables,
the important point to notice is that the resummation of NGLs is included as a
factor multiplying the single--gluon Sudakov form factor,
$\Sigma_{P}$,~\cite{Dasgupta:2001sh}
\begin{equation}\label{resum_tot}
\Sigma\left(\tauo, \Eo\right) = \Sigma_{P}(\tauo,\Eo)\; S\left( t\right),
\end{equation}
In this section, we first consider resummation of $\tauo$ distribution in events
where the final state jets are defined in the anti--$\kt$ algorithm and, second,
discuss the potential changes to the resummed result when the jets are defined
in the C--A algorithm instead.

\subsection{Resummation with anti--$\kt$ algorithm}\label{subsec.resummation_QCD}

As stated in the introduction and proved in sec.~\ref{sec.fixed_order_1}, the
$\tauo$ observable is simply the sum of the invariant masses of the two
highest--energy (or highest--$\pt$ for hadron colliders) jets. Therefore the
$\tauo$ resummed Sudakov form factor is just double that computed
in~\cite{Banfi:2010pa}, for a single jet mass. That is, up to NLL level we have
\begin{equation}\label{resum_tot_akt}
\Sigma_{P}(\tauo,\Eo) = \frac{\exp\left[-2\left(\R_{\tauo}(\tauo) + \gamma_{E}
\R'_{\tauo}(\tauo) \right)\right]}{\Gamma\left(1+2\,\R'_{\tauo}(\tauo)\right)}\;
\exp\left[ - \R_{\Eo}(\Eo)\right] .
\end{equation}
The full derivation of~\eqref{resum_tot_akt} as well as the resultant
expressions of the various radiators are presented in the small jet--radius
limit in Ref.~\cite{Banfi:2010pa}. To restore the full $\Rs$ dependence we make
the replacement $R^{2}/\rho \mapsto \Rs/(\tauo (1-\Rs))$ such that when expanded
eq.~\eqref{resum_tot_akt} reproduces at $\Or(\as)$ the LO
distribution~\eqref{R1_full-b}.

To account for the NGLs at all--orders in $\as$, it is necessary to consider an
arbitrary ensemble of energy--ordered, soft wide--angle gluons that coherently
radiate a softest gluon into the vetoed region of phase
space~\cite{Dasgupta:2001sh}~\footnote{In our case the vetoed region is the jet region. Due to symmetry, we can choose one jet region and multiply the final answer by a factor of two.}. The analytical resummation of NGLs is then plague with mathematical problems coming from geometric and colour structure of the gluon ensemble. Two methods have been developed to address this issue: A numerical Monte Carlo evaluation~\cite{Dasgupta:2001sh,Dasgupta:2002bw} and a non--linear evolution equation that resums single logs (SL) at all--orders~\cite{Banfi:2002hw}. Both methods are only valid in the large--$\Nc$ limit. In the latter limit and for small values of the jet--radius $\Rs$, we
argued in~\cite{Banfi:2010pa} that the form of $S(t)$ should be identical to
that found in the hemisphere jet mass case~\cite{Dasgupta:2001sh}. Since in the
present paper we are not confined to the small $\Rs$ limit, we need to modify
and re--run the Monte Carlo algorithm, presented in~\cite{Dasgupta:2001sh}, for
medium and large values of the jet--radius should we seek to resum the $\tauo$
NGLs distribution. The latter task is, however, beyond the scope of this paper.
Here, we are only aiming at comparing the analytical results with fixed--order
NLO program \texttt{EVENT2}. It suffices in this case to simply exponentiate the
first NGLs term in eq.~\eqref{S2_akt_tot},
\begin{equation}\label{resum_NGLs_akt}
S(t) = \exp\left(S_{2}\; t^{2}\right).
\end{equation}

The distribution~\eqref{resum_tot} is of the generic form given in eq.~\eqref{resum_gen_struct}. Explicitly, it reads
\begin{equation}\label{resum-form_QCD-b}
\Sigma(\tauo, \Eo) = \left(1+\sum_{k=1}^{\infty} C_{k}
\left(\frac{\as}{2\pi}\right)^{k} \right)
\exp\left[\sum_{n=1}^{\infty}\sum_{m=0}^{n+1} G_{nm}
\left(\frac{\as}{2\pi}\right)^{n} \widetilde{L}^{m} \right] +
D_{\mathrm{fin}}(\tauo),
\end{equation}
where $C_{k}$ is the $k^{th}$ loop--constant, $\widetilde{L} = \ln(1/\tauo)$ and $D_{\mathrm{fin}} \equiv D$, which vanishes in the limit $\tauo \rightarrow 0$. In order to determine the coefficients $G_{nm}$ at NLO and up to NLL,we need to expand the radiators, as well as the $\Gamma$ function, in
eq.~\eqref{resum_tot_akt} up to second order in the fixed coupling $\as =
\as(Q)$. The results are presented in appendix.~\ref{app.coeff_in_expansion}.
Although we have provided the NNLL coefficient, $G_{21}$ in
eq.~\eqref{G_nm-QCD}, we do not claim that it is under control. Nonetheless, it
does capture all $\Rs$--dependent terms~\footnote{as can be seen from comparison
to the SCET result~\eqref{app.tw_in_SCET}, which only contains the primary
emission piece and is valid to NNLL.}. The missing terms from $G_{21}$ include:
a) coefficients of $\widetilde{L}$ which are independent of $\ln(\Rs/(1-\Rs))$
for all colour channels. These can be borrowed from thrust
distribution~\cite{Kelley:2011ng,Monni:2011gb,Hornig:2011iu}. b) Although $G_{21}$ has a subleading NGLs term in the $\CF\CA$ colour channel, which comes solely from the expansion of $t$ (eq.~\eqref{S2_akt_tot}), the full expression in this channel as well as in the $\CF\TF\nf$ channel is still missing. To properly compute the latter, one has to extend both the matrix--element~\eqref{W_2} and the phase space to include hard emission. Such a task will be considered elsewhere. It is worthwhile to mention that full subleading NGLs have recently been computed analytically within SCET framework for the hemisphere mass variable~\cite{Kelley:2011ng,Hornig:2011iu}~\footnote{Recall that for the
leading NGLs the corresponding coefficient for the hemisphere mass distribution
corresponds to setting $\Rs = 0$ in $S_{2,a}$~\eqref{S2_akt_a1}.}. The two--loop
constant $C_{2}$ has also been computed for the latter variable as well as the thrust~\cite{Kelley:2011ng,Monni:2011gb}.

To make contact with SCET calculations, we provide in
appendix~\ref{app.tw_in_SCET} the full formula of the Sudakov form factor for
the $\tauo$ primary distribution including determination of the $G_{nm}$
coefficients in SCET. 

Next we comment on the form of resummation when final state jets are defined in
the C--A algorithm.

\subsection{Resummation with C--A algorithm}\label{subsec.resum_CA}

With regard to primary emission  piece, resumming logs induced by clustering is
a cumbersome but doable task. It has been performed, for example,
in~\cite{Delenda:2006nf} for interjet energy flow distribution where final state
jets are defined in the inclusive $\kt$ algorithm. The final result of the
resummed radiator was written as an expansion in the jet--radius and the first
four terms were determined. For secondary emissions, the resummation of NGLs has
only been possible numerically and in the large--$\Nc$ limit. It has again been
carried out for the above mentioned  energy flow distribution
in~\cite{Appleby:2002ke}. We expect that analogous, to the interjet energy flow,
analytical treatment and numerical evaluation can be achieved for the
resummation of CLs and NGLs, respectively, for the $\tauo$ variable. We postpone
this work to future publications.

For the sake of comparing to \texttt{EVENT2}, it is sufficient to simply
exponentiate the fixed--order terms $S^{\ca}_{2}$ and $C_{2}^{P}$, just as we
did with the anti--$\kt$ algorithm case. Due to the fact that logarithmic
contributions induced by clustering arise mainly from soft wide--angle gluons,
we expect them -clustering--induced logs- to factorise from the primary form
factor at all--orders. Therefore, the resummed distribution, whereby clustering
is imposed on the final state, may be written as
\begin{equation}\label{resum_tot_CA}
\Sigma(\tauo,\Eo) = \Sigma_{P}(\tauo, \Eo) \,S^{\ca}\left(t\right)
C^{P}\left(t_{p}\right),
\end{equation}
where $S^{\ca}$ is of the form~\eqref{resum_NGLs_akt} with $S_{2}$ replaced by
$S_{2}^{\ca}$ and, in analogy with the NGLs factor, the CLs factor reads
\begin{equation}\label{resum_C_P_CA}
C^{P}\left(t_{p}\right) = \exp\left(C_{2}^{P}\,t_{p}^{2}\right),\;\;\; t_{p} =
\int^{Q/2}_{Q\tauo/2 \Rs} \frac{\d\kt}{\kt} \frac{\as(\kt)}{2\pi}.
\end{equation}

\begin{figure}[t]
\centering
  \includegraphics[width=7.5cm]{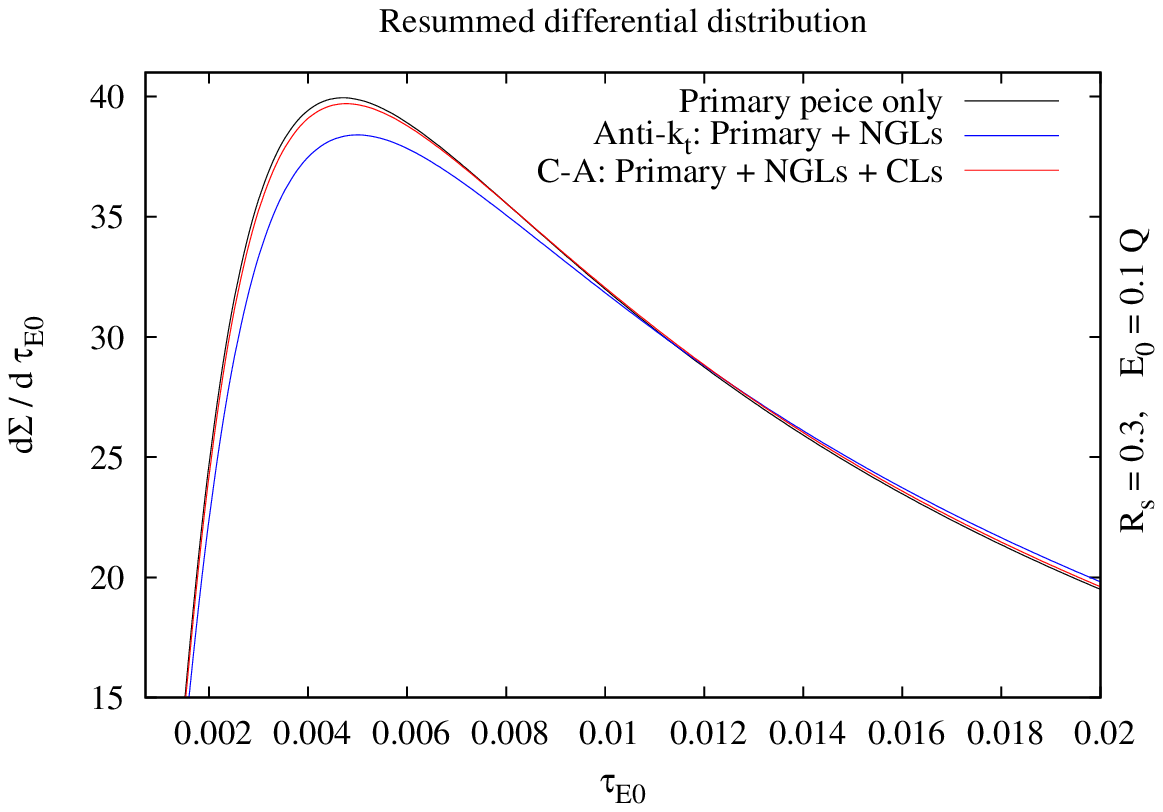}
  \includegraphics[width=7.5cm]{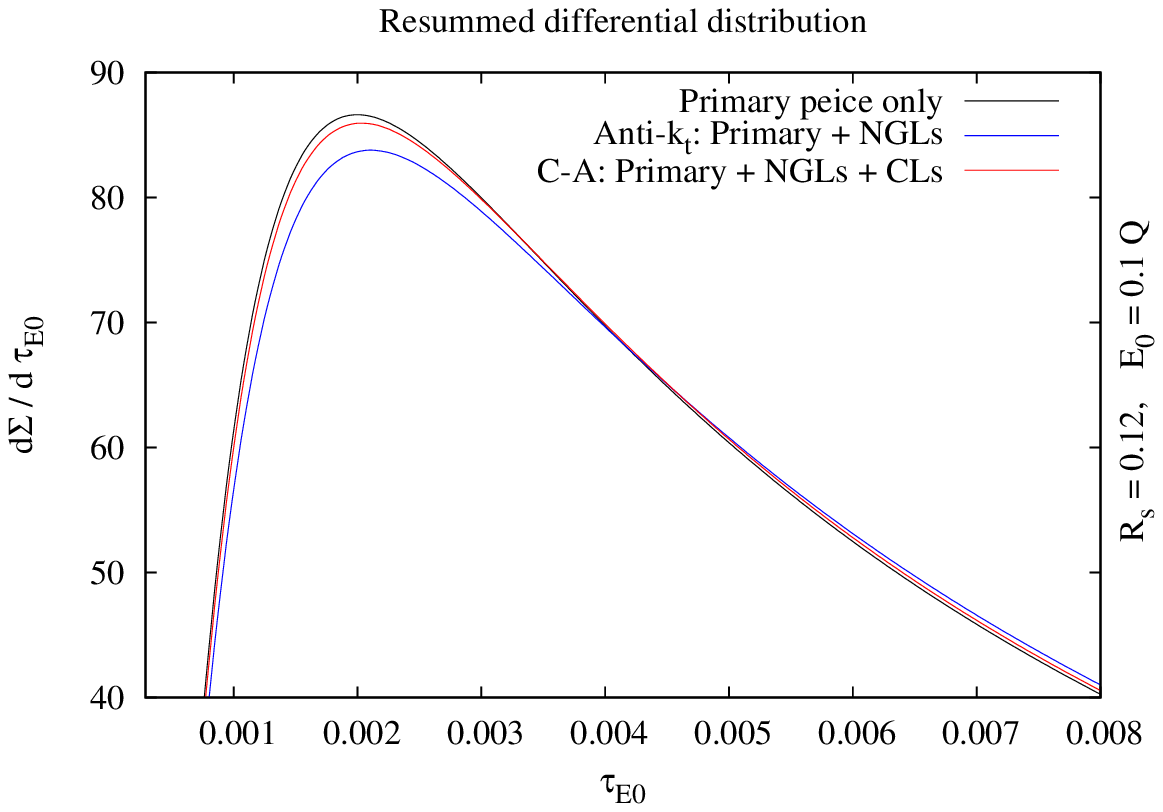}	\includegraphics[width=7.5cm]{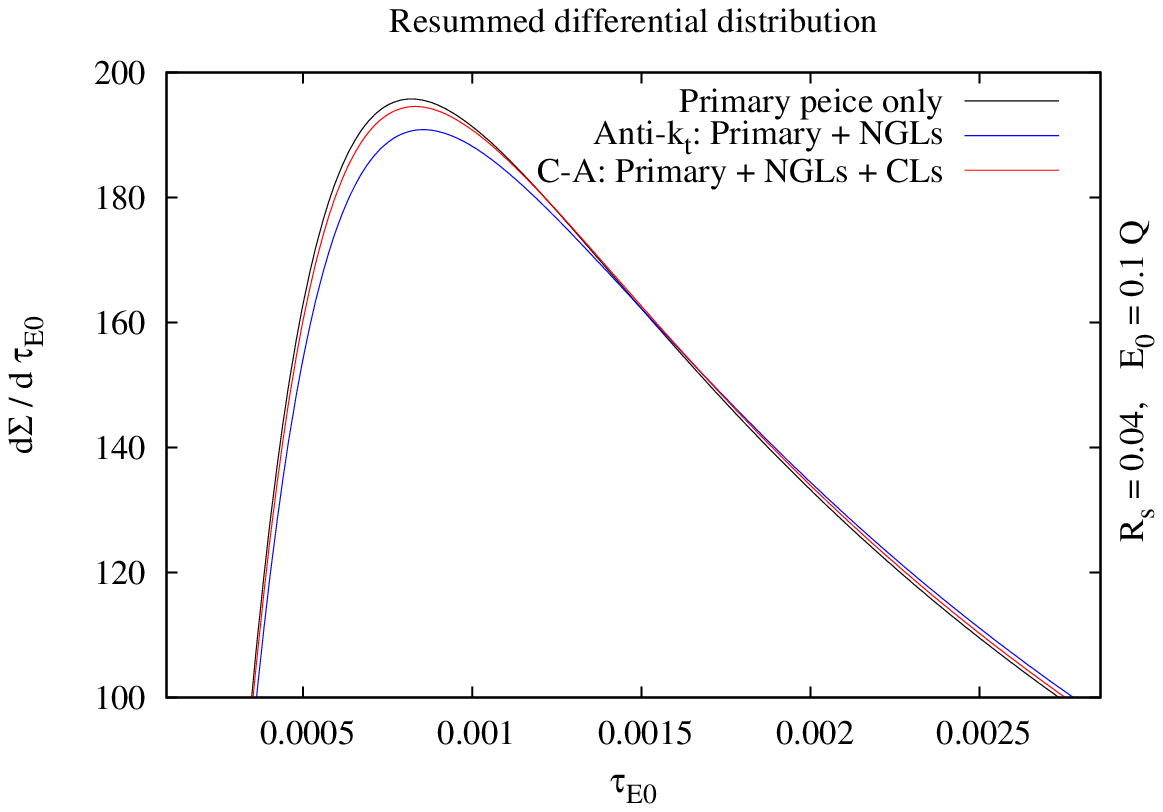}		\includegraphics[width=7.5cm]{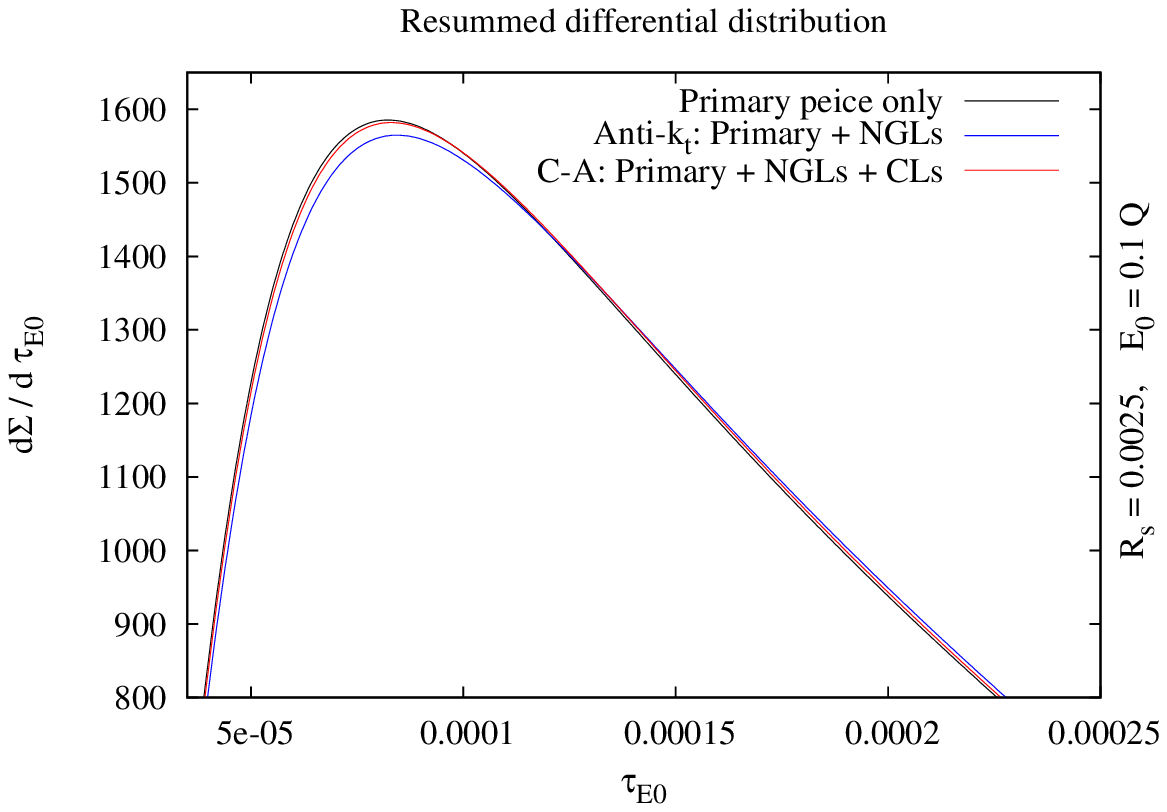}				\caption{Comparison of analytical resummed differential distribution
$\d\Sigma/\d\tauo$ where: only primary term included~\eqref{resum_tot_akt},
primary and NGLs factor included in the anti--$\kt$ algorithm~\eqref{resum_tot_akt} and primary $+$ NGLs $+$ CLs factors included~\eqref{resum_tot_CA}. The plots are shown for various jet--radii with a jet veto $\Eo = 0.1 Q$. The coupling is taken at the $Z$ mass to be $\as(M_{Z}) \simeq 0.118$. The plots are only meant to give a rough estimate of the effects of NGLs in non--clustered as well as clustered final states.}
\label{fig.resummed_CA_akt}
\end{figure}

In fig.~\ref{fig.resummed_CA_akt} we plot the resummed differential
distributions, $\d \Sigma(\tauo,\Eo)/\d\tauo = (1/\so)\d\sigma/\d\tauo$, computed
from eq.~\eqref{resum_tot_akt} for the anti--$\kt$ algorithm and from
eq~\eqref{resum_tot_CA} for the C--A algorithm at different values of $\Rs$. The
dependence on $\Eo$ has been discussed in~\cite{Banfi:2010pa} where the
all--orders NGLs resummed expression was employed. There are several points to note. Firstly the effect of NGLs is a suppression of the total cross--section relative to the primary result. This suppression is diminished by decreasing the value of $\Rs$. For example, at $\Eo=0.1 Q$ the Sudakov peak is reduced due to NGLs by about $4.02\%, 3.42\%, 2.62\%$ and $1.35\%$ for $\Rs = 0.30, 0.12, 0.04$ and $0.0025$ (equivalent to $R = 1.1, 0.7, 0.4$ and $0.1$) respectively. These values are only meant to give an idea of the effect of varying the jet--radius parameter on both NGLs and CLs corrections to the total cross--section, since we are only working with an approximation of the latter and not the full all--orders result. It has been shown in~\cite{Appleby:2002ke}, for the interjet energy distribution, that the NGLs resummed factor $S(t)$ at all--orders is much smaller (thus larger suppression of primary--only result) and of different shape, as a function of $t$, to the fixed--order exponentiated result.

Secondly the effect of clustering is reducing the phenomenological significance
of NGLs. This reduction becomes larger, hence the NGLs suppression on the
Sudakov peak becomes smaller, as one moves towards smaller values of $\Rs$. For the same jet veto $\Eo = 0.1 Q$, the Sudakov peak is reduced by $0.62\%$~\footnote{The discrepancy at $\Rs = 0.30$ is due to the fact that we have employed the small angles approximation in the C--A calculations.}$, 0.80\%, 0.63\%$ and $0.22\%$ for $\Rs = 0.30, 0.12, 0.04$ and $0.0025$ respectively (values are only an estimate of the impact of clustering). Comparing to the anti--$\kt$ case, we see that the effect of NGLs has been reduced by more than $70\%$ for $\Rs= 0.12, 0.04$ and $\Rs = 0.0025$. This observation suggests that instead of resumming NGLs, which is a daunting task even numerically, one
should, perhaps, attempt at eliminating them at each order through requiring final state clustering and looking for the optimal value of the jet--radius, and may be the jet veto too, such that non--global corrections are wiped out. In our rough approximation, we find that NGLs are completely eliminated, leaving only the primary Sudakov form factor, at $\Rs \lesssim 3\times 10^{-5}$ (equivalent to $R \lesssim 0.01$). Although this value is very small and not of any practical significance, including the full all--orders resummed results for both NGLs and CLs might result in practically larger values. Whether this is indeed the case remains to be investigated. If it turns out that the optimal radius is relatively large, $0.04 \lesssim \Rs\; (0.4 \lesssim R)$, then final state clustering will be the key to solve the NGLs subtlety of non--global observables.

In the next section, we compare our analytical calculations to \texttt{EVENT2}.
In particular, we focus on establishing the presence of NGLs and CLs in the
$\tauo$ distribution at NLO.

\section{Numerical results} \label{sec.numerical_results}

The $\tauo$ numerical distribution has been computed using the fixed--order NLO
QCD program \texttt{EVENT2}. The program implements the Catani--Seymour
subtraction formalism for NLO corrections to two-- and three--jet events
observables in $\ee$ annihilation. Final state partons have been clustered into
jets using the \texttt{FastJet} library~\cite{Cacciari:2011ma}. The latter provides an implementation
of the longitudinally invariant $\kt$, Cambridge--Aachen (CA) and anti--$\kt$
jet finders along with many others. Cone algorithms such as SISCone~\cite{Salam:2007xv} are also implemented as plugins for the package. It should be noted that the $\ee$
version of the aforementioned algorithms employs the following clustering
condition for a pair of partons $(ij)$ 
\begin{equation}\label{clust_cond_FJ}
    1-\cos\theta_{ij} < 1-\cos(\widetilde{R}),
\end{equation}
where $\widetilde{R}$ is the jet--radius parameter used in
\texttt{FastJet}~\footnote{In \texttt{FastJet}'s manual $\widetilde{R}$ is allowed
to go up to $\pi$. Since we are interested in two--jet events the jet size
cannot be wider than a hemisphere. Thus we restrict $\tilde{R}$ to be less than
$\pi/2$.}. Compared to eqs~\eqref{clust_cond} and~\eqref{R_Rs_rel}, $\widetilde{R} =
\cos^{-1}(1-2\Rs)$. The exact numerical distributions $(1/\so)(\d\sigma_{e}/\d
L)$, with $L = -\widetilde{L} = \ln(\tauo)$,.
 for the three colour channels, $\CF^{2}, \CF\CA$ and $\CF\TF\nf$, have been
obtained with $10^{11}$ events in the bin range $0>L>-14$. We have used four
values for the jet--radius: $\Rs = 0.50, \Rs = 0.30, \Rs=0.12$ and $\Rs = 0.04$, with
an energy veto $\Eo = 0.01\,Q$. Standard deviations on individual bins
range from $10^{-4}\% $ to $ 10^{-2}\%$.

We plot the difference between the numerical and analytical distributions at
both LO and NLO,
\begin{equation}\label{remainder}
 r(L) = \frac{\d\sigma_{e}}{\so\d \,L} - \frac{\d\sigma_{r,2}}{\so\d L},
\end{equation}
where $\d\sigma_{r,2}/\so\d L$ is given in eq.~\eqref{resum_expanded-sig-diff}.
Recall that at small values of the jet shape, $\tauo$, the finite remainder
function $D_{\mathrm{fin}}(\tauo)$ is vanishingly small and will thus be
ignored. For the case where the jet shape is global ($\Rs = 0.50$ and the
threshold thrust reduces to thrust), we expect a full cancellation of singular
terms and thus $r$ should be a constant line corresponding to the NNLL
coefficient ($H_{21}$ in eq.~\eqref{H_nm_coeffs}). For $\Rs < 0.5$, the jet
shape is non--global and we expect $r$ to have a slope if NGLs contribution is
excluded. If our analytical calculations of the NGLs' coefficient, both for
anti--$\kt$ and C--A algorithms, are correct then upon adding the latter to
$H_{22}$ the slope should vanish and $r$ becomes flat signalling a complete
cancellation of terms up to NLL level. Similar behaviour should be seen with the
CLs' coefficient $C_{2}^{P}$ for the C--A algorithm case. Considering
figs.~\ref{fig.LO} - \ref{fig.CA_CFTF}, we make the following observations:
\begin{itemize}
  \item At LO, the distribution is independent of the jet definition. From
eq.~\eqref{H_nm_coeffs} we have
\begin{equation}
H_{11} = - 3 + 4 L_{\Rs} = -4, -8.51, -14.75, -20.95 ;\;\;\; \mathrm{for}\; \Rs
= 0.5, 0.3, 0.12, 0.04.
\end{equation}
Compared to the numerical results shown in fig.~\ref{fig.LO} we see a complete
agreement. The cut--off in fig.~\ref{fig.LO} is due to the fact that at LO
$\tauo < \Rs/(1+\Rs)$ (eq.~\eqref{R1_full-b}).
\begin{figure}[t]
	\centering
	\includegraphics[width=7.4cm]{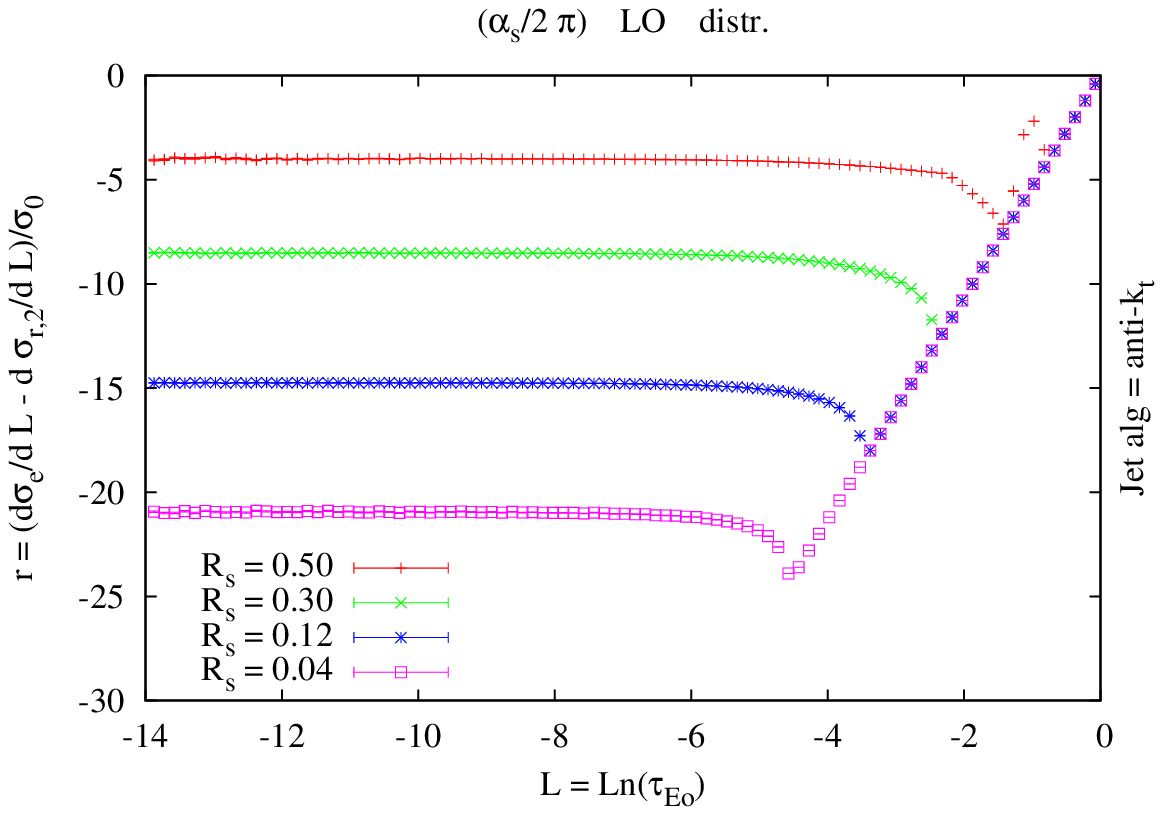}
        \includegraphics[width=7.4cm]{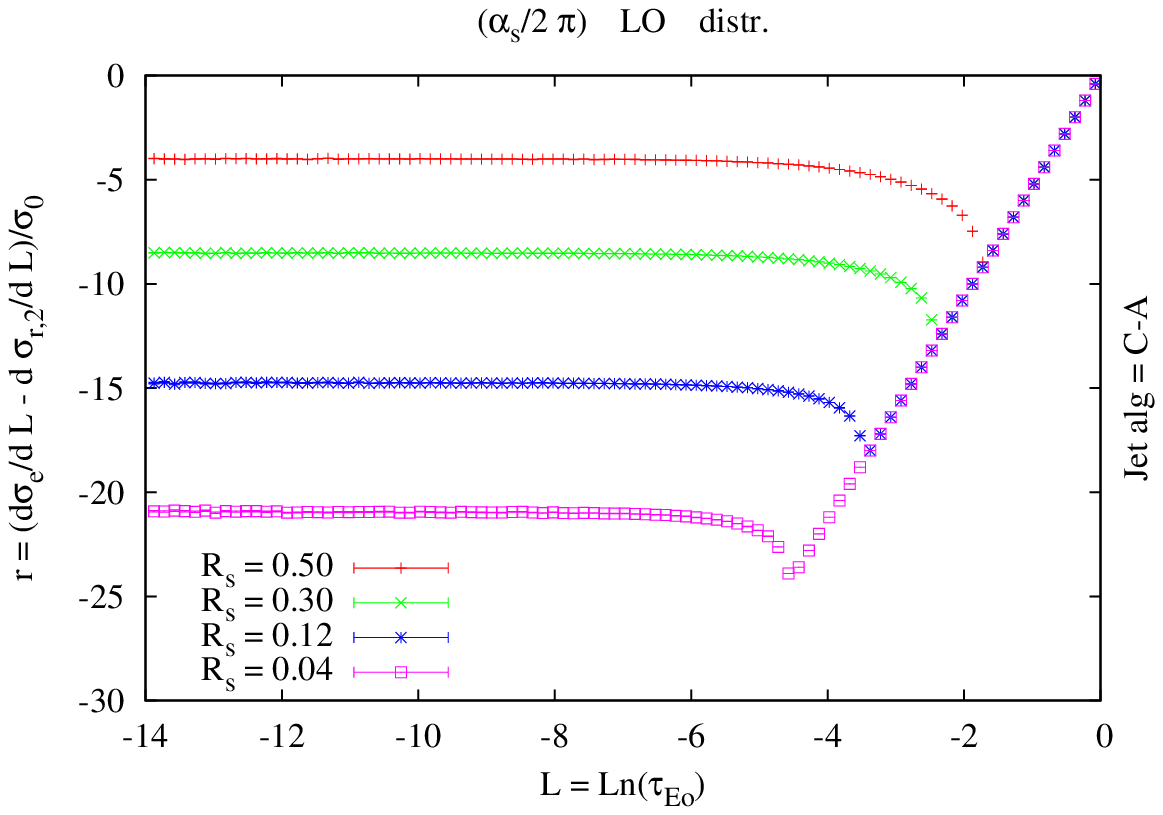}
	\caption{The difference between \texttt{EVENT2} and $\tauo$ LO
distribution for various jet radii in both anti--$\kt$ (left) and CA (right)
algorithms.}
	\label{fig.LO}
\end{figure}
   \item For the NLO distribution in the anti--$\kt$ algorithm,
fig.~\ref{fig.AKT_CFCA} (left) illustrates the existence of NGLs. The flatness of the
$r(L)$ curve, in fig.~\ref{fig.AKT_CFCA} (right), at $L$ below about $-9$ indicates a complete cancellation up to
single log level.
\begin{figure}[t]
	\centering
	\includegraphics[width=7.4cm]{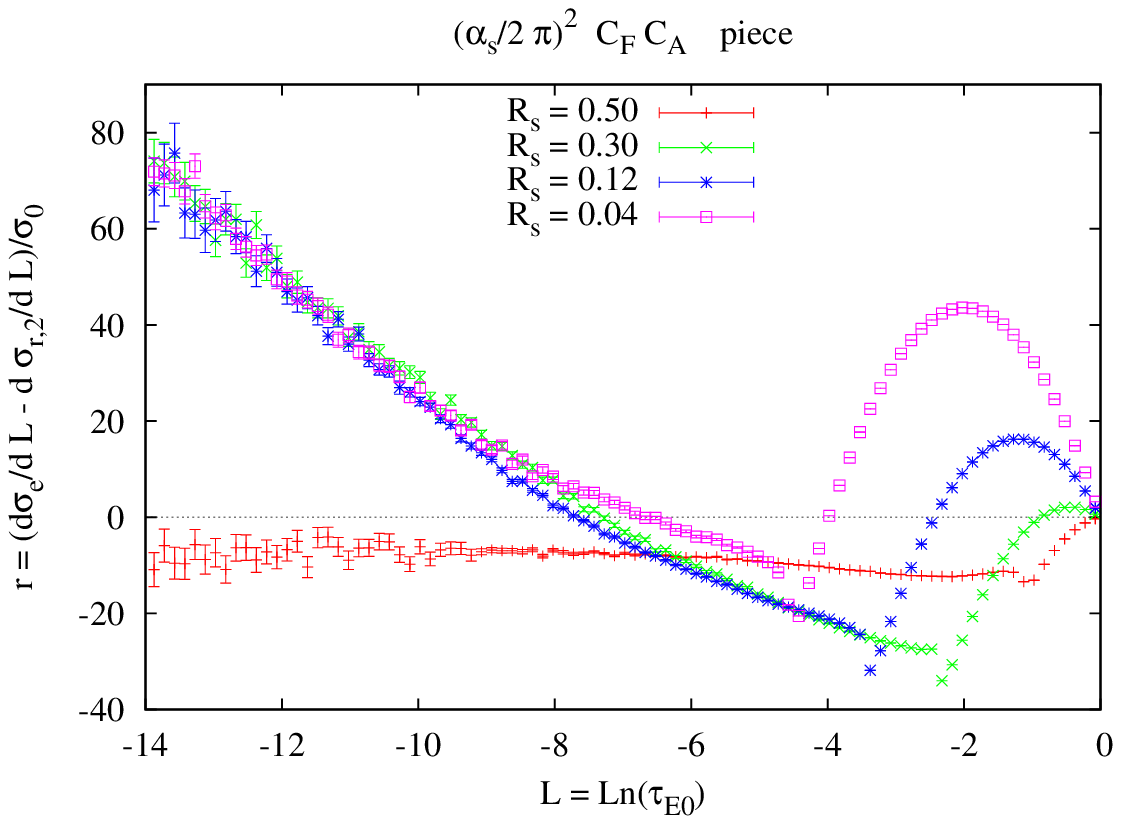}
        \includegraphics[width=7.4cm]{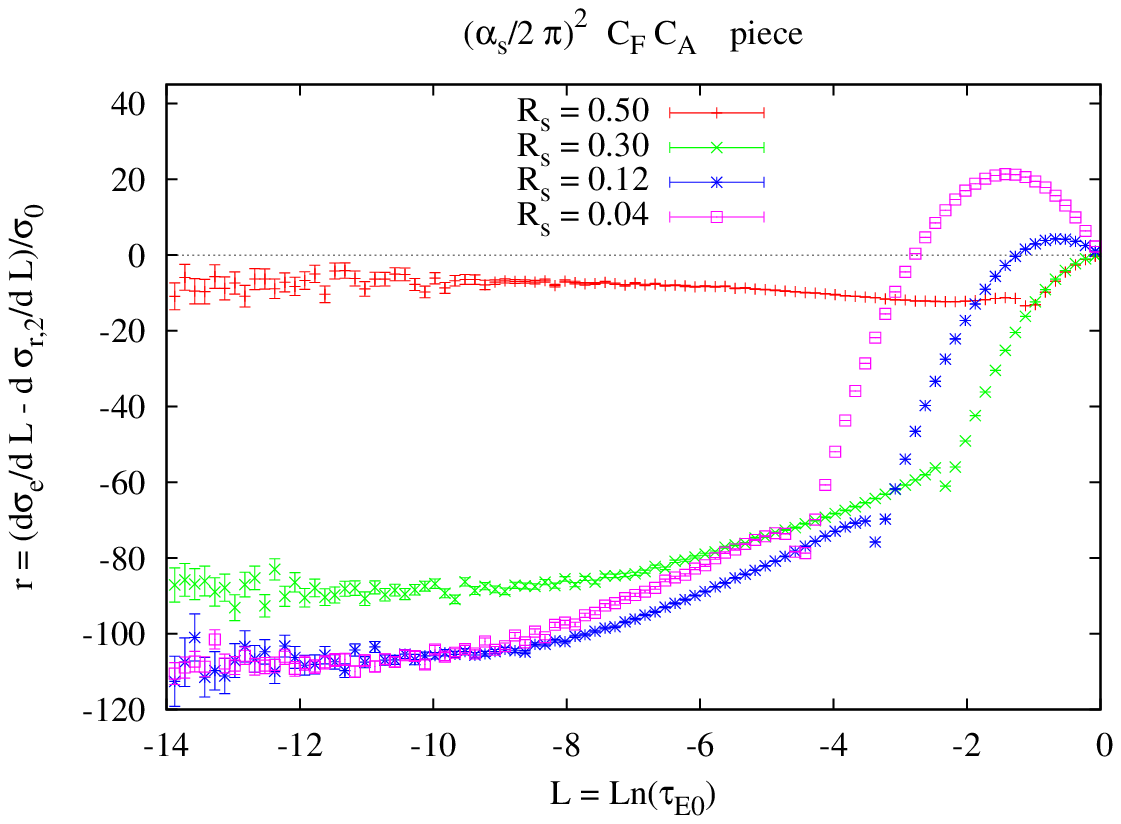}
	\caption{The $\CF\CA$ part of the difference between \texttt{EVENT2} and
(left) $\tauo$ primary (global) distribution and (right) $\tauo$ distribution
including NGLs for various jet radii in anti--$\kt$ algorithm.}
	\label{fig.AKT_CFCA}
\end{figure}
\begin{table}[t]
\centering
\begin{tabular}[c]{cc|c|c||c|c||c|c|}
 \cline{3-8}
  & & \multicolumn{2}{|c|}{$\CF\TF\nf$ piece of} & 
      \multicolumn{2}{|c|}{$\CF\CA$ piece of}    &  
      \multicolumn{2}{|c|}{$\CF^{2}$ piece of} 
  \\ \cline{3-8}
 $\Rs$ & Jet alg & $H_{21}^{\mathrm{num}}$ & $H_{21}^{\mathrm{analyt}}$ & 
      $H_{21}^{\mathrm{num}}$ & $H_{21}^{\mathrm{analyt}}$ &
      $H_{21}^{\mathrm{num}}$ & $H_{21}^{\mathrm{analyt}}$
  \\ \cline{1-8}
  \multicolumn{1}{|c|}{\multirow{2}{*}{$0.50$}}  &
  \multicolumn{1}{|c|}{anti--$\kt$} & $5.20 \pm 0.14$    & $5.00$ &
  \multicolumn{1}{|c|}{$ -7.26 \pm 0.22 $}  &   $ -7.04$ &    
  \multicolumn{1}{|c|}{$ -12.40  \pm 0.57 $}  & $ -12.68 $       
\\ \cline{2-3} \cline{5-5} \cline {7-7}
  \multicolumn{1}{|c|}{}                                 &
  \multicolumn{1}{|c|}{C--A}          & $4.99 \pm 0.19$  &        &
  \multicolumn{1}{|c|}{$ -6.97 \pm 0.11 $} &             &
  \multicolumn{1}{|c|}{$ -12.99 \pm 0.67 $}           &          
\\ \cline{1-8} 
  \multicolumn{1}{|c|}{\multirow{2}{*}{$0.30$}}  &
  \multicolumn{1}{|c|}{anti--$\kt$} &$ 13.92 \pm 0.15$   & $7.80$ &
  \multicolumn{1}{|c|}{$ -88.35 \pm 0.26 $}  & $ -11.45$ &  
  \multicolumn{1}{|c|}{$ 62.73 \pm 0.51 $}   & $ 62.20 $ 
\\ \cline{2-3} \cline{5-5} \cline{7-7}
  \multicolumn{1}{|c|}{}                                 &
  \multicolumn{1}{|c|}{C--A}          & $11.24 \pm 0.08$  &        &
  \multicolumn{1}{|c|}{$ -48.46 \pm 0.31$}  &            &
  \multicolumn{1}{|c|}{$ 76.38 \pm 0.50$}  &  
\\ \cline{1-8}
  \multicolumn{1}{|c|}{\multirow{2}{*}{$ 0.12$}}  &
  \multicolumn{1}{|c|}{anti--$\kt$} &$ 15.75 \pm 0.12$   & $8.54$ &
  \multicolumn{1}{|c|}{$ -106.80 \pm 0.34$}  & $ -8.89$  &
  \multicolumn{1}{|c|}{$ 384.04 \pm 0.90$}   & $ 385.27 $  
\\ \cline{2-3} \cline{5-5} \cline{7-7}
  \multicolumn{1}{|c|}{}                                 &
  \multicolumn{1}{|c|}{C--A}          &$ 13.65 \pm 0.08$ &        &
  \multicolumn{1}{|c|}{$ -51.46 \pm 0.35$}  &            &
  \multicolumn{1}{|c|}{$ 405.48 \pm 0.57$}  & 
\\ \cline{1-8}  
  \multicolumn{1}{|c|}{\multirow{2}{*}{$ 0.04$}}  &
  \multicolumn{1}{|c|}{anti--$\kt$} &$ 12.86 \pm 0.19$  & $5.66$  &
  \multicolumn{1}{|c|}{$ -107.67 \pm 0.28$}  & $ 3.59$ &
  \multicolumn{1}{|c|}{$ 1017.93 \pm 0.53$}  & $ 1022.43$   
\\ \cline{2-3} \cline{5-5} \cline{7-7}
  \multicolumn{1}{|c|}{}                               &
  \multicolumn{1}{|c|}{C--A}     & $ 10.60 \pm 0.23$   &          &
  \multicolumn{1}{|c|}{$ -45.67 \pm 0.23$}  &          &
  \multicolumn{1}{|c|}{$ 1044.33\pm 1.24$}  & 
\\ \cline{1-8}  
 \hline
\end{tabular}
\caption{$H_{21}$ numerical vs analytical values for all three colour pieces. The
numerical values were obtained through fitting the \emph{flat} curve $r(\widetilde{L})$ with the function (see eq.~\eqref{singular_terms_gen}) $H_{21}^{\mathrel{num}} + e^{-\widetilde{L}}(B + C\,e^{-\widetilde{L}})$. The analytical values, $H_{21}^{\mathrm{analyt}}$, are taken from eqs.~\eqref{H_nm_coeffs} and~\eqref{G_nm_SCET} which only include the primary emission contribution with neither non--global nor clustering terms. $\Rs=0.50$ corresponds
to the global case and we expect the analytical and numerical values to be the
same.}\label{tab:H21_akt}
\end{table}
The $\CF^{2}$ and $\CF\TF\nf$ pieces are shown in fig.~\ref{fig.AKT_CF2-CFTF}. In table~\ref{tab:H21_akt} we provide both numerical and analytical,
taken from SCET calculations~\eqref{G_nm_SCET}, values of the NNLL coefficient,
$H_{21}$, at the considered $\Rs$ values for the three colour channels. It is evident from the table that there are subleading $\Rs$--dependent NGLs for both $\CF\CA$ and $\CF\TF\nf$ channels. Such logs have been analytically computed in~\cite{Kelley:2011ng} for the hemisphere jet mass. Our numerical results show that they are also present for finite--size jets~\footnote{as would be anticipated, since the finite--size jet mass is an extension to the hemisphere mass.}. The primary $\CF^{2}$ channel is free from such subleading NGLs as numerical and analytical values of $H_{21}$ in the anti--$\kt$ coincide.

Notice that while the $x$--axis in all figures shown in this section corresponds to $\ln(\tauo) = \log(\tauo)$, i.e, the natural logarithm of the jet shape, that of~\cite{Hornig:2011tg} corresponds to the logarithm of base $10$, $\log_{10}(\rho) \sim \log(\rho)/2, \; \rho \equiv \tauo$, of the jet shape. Given this, fig.~\ref{fig.AKT_CFCA} above is equivalent to fig.~$7$ of~\cite{Hornig:2011tg}. Neither $\CF^{2}, \CF\TF\nf$ plots nor subleading NGLs were considered in~\cite{Hornig:2011tg}.
\begin{figure}[t]
	\centering
	\includegraphics[width=7.5cm]{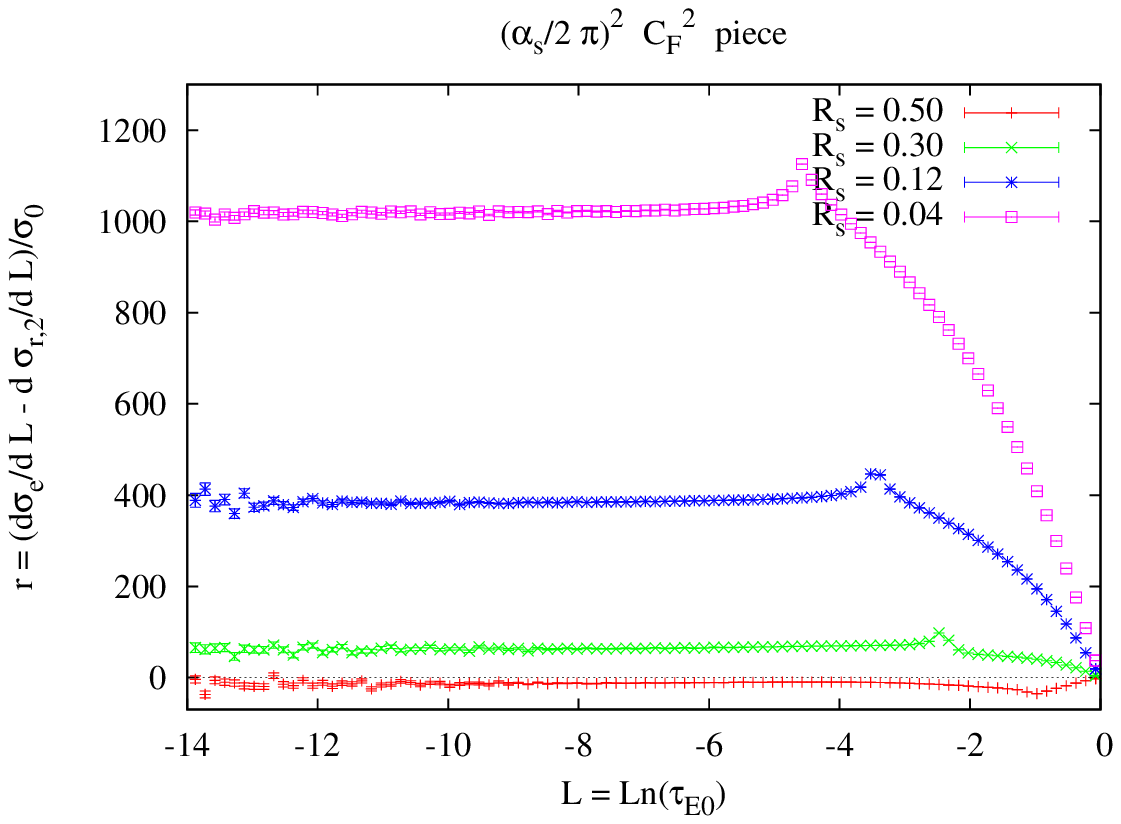}
        \includegraphics[width=7.5cm]{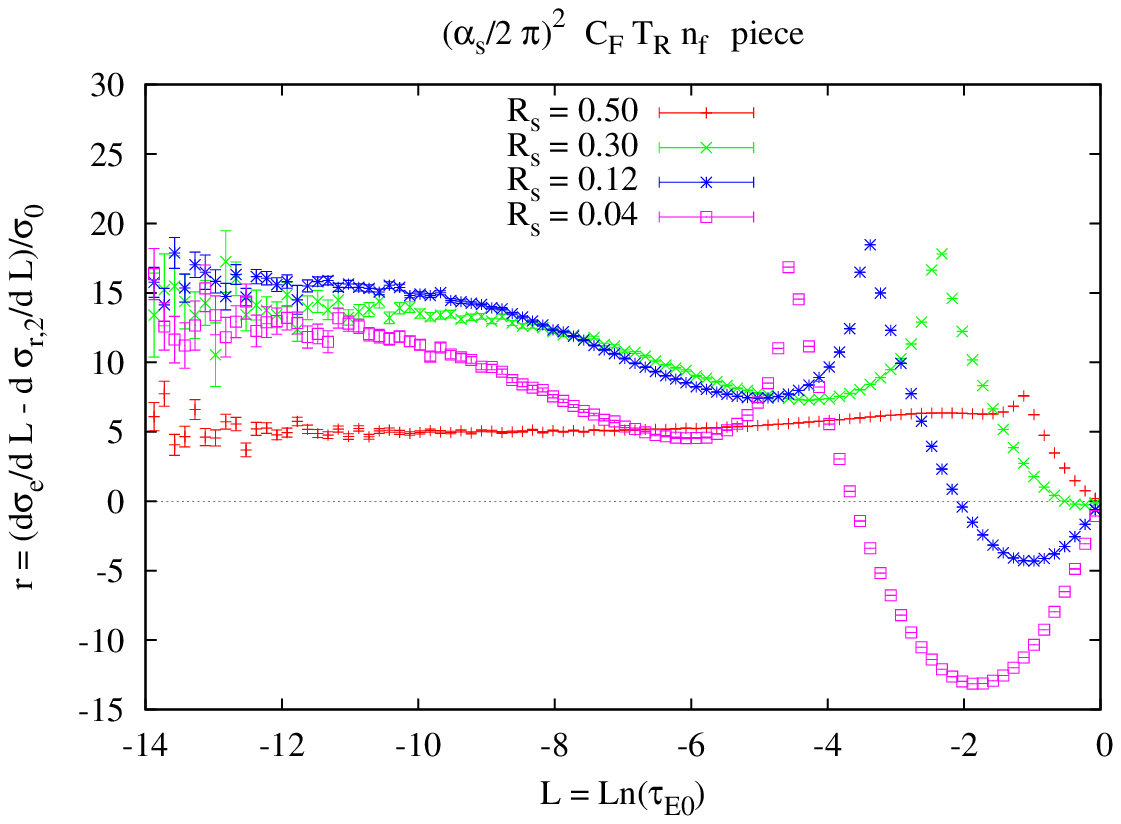}
	\caption{The (left) $\CF^{2}$  and (right) $\CF\TF\nf$ piece of the
difference between \texttt{EVENT2} and $\tauo$ distribution for various jet
radii in the anti--$\kt$ algorithm.}
	\label{fig.AKT_CF2-CFTF}
\end{figure}
  \item The asymptotic region, i.e, the region where large logs are expected to
dominate over non--logarithmic contributions, corresponds to $L$ less than about
$-9$ (for figs.~\ref{fig.AKT_CFCA}, $\CF^{2}$ piece in
fig~\ref{fig.AKT_CF2-CFTF} and may even be less for the $\CF\TF\nf$ piece in
fig.~\ref{fig.AKT_CF2-CFTF}) and seems to decrease further as $\Rs$ becomes smaller. A
similar effect is seen in the thrust distribution, fig.~\ref{fig.thrust_NLO},
where the numerical distribution has been obtained using the full
definition~\eqref{thrust_def}.
 \begin{figure}[t]
	\centering
	\includegraphics[width=7.5cm]{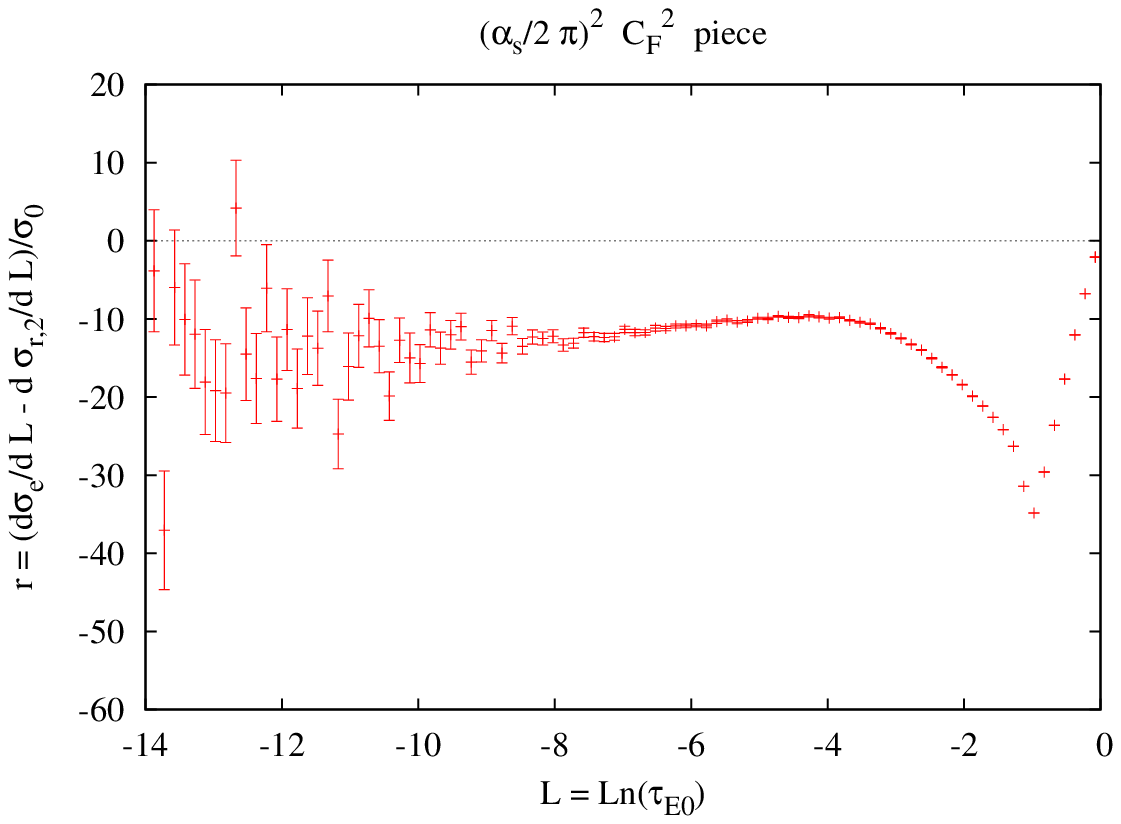}
	\includegraphics[width=7.5cm]{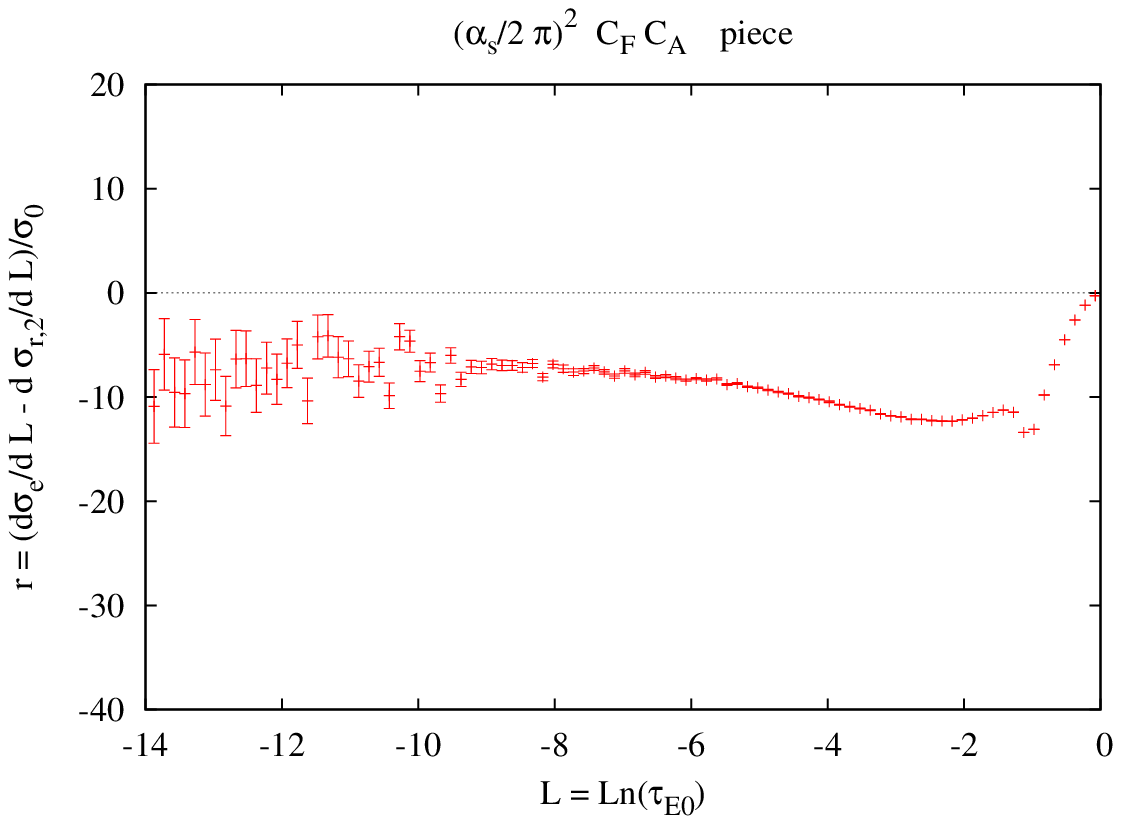}
	\includegraphics[width=7.5cm]{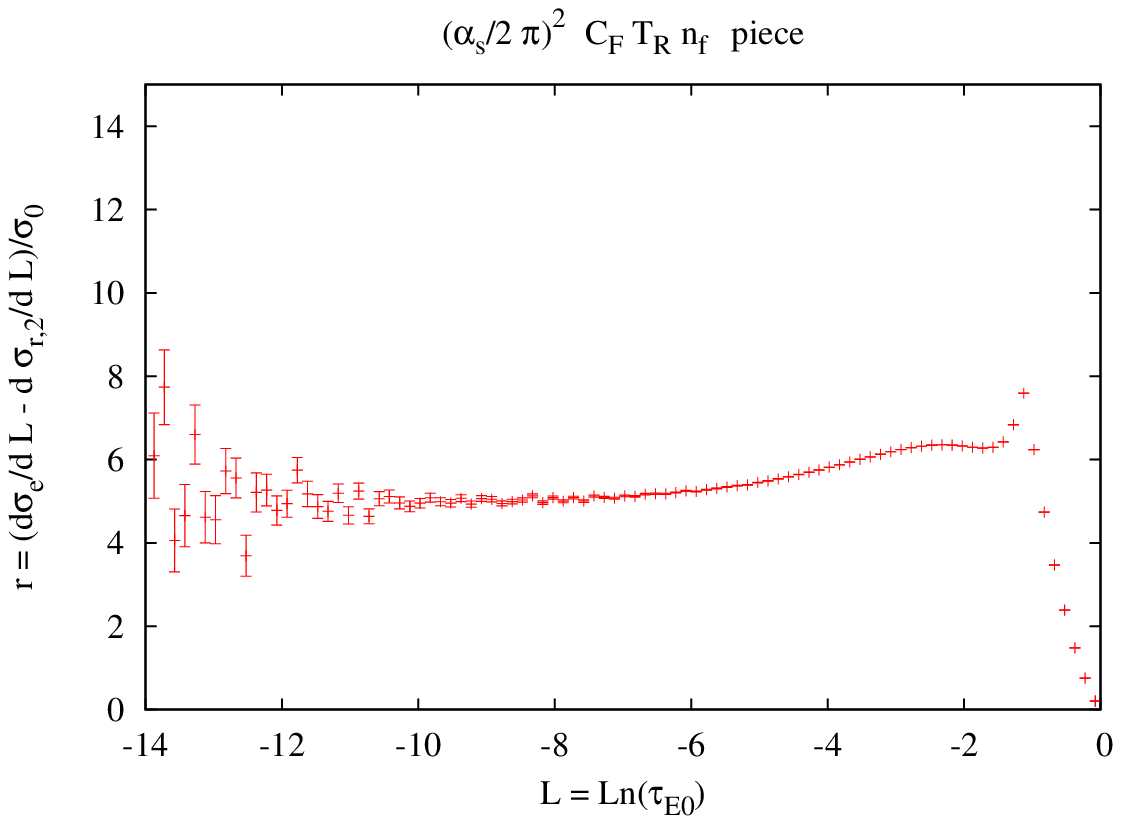}
	\caption{The various colour pieces of the difference between
\texttt{EVENT2} and thrust distribution using the full
definition~\eqref{thrust_def}. The pQCD resummed analytical expression for
thrust distribution can be found in, for example,~\cite{QCD_collider}.}
	\label{fig.thrust_NLO}
\end{figure}
  \item Considering the clustering case with C--A algorithm,
figs.~\ref{fig.CA_CF2} and~\ref{fig.CA_CF2_zoom} illustrate the presence of CLs
in the $\CF^{2}$ channel. Clearly, the addition of CLs makes the remainder $r$
flat in the region $L \lesssim -9$. To strengthen this observation even more, we
plot in fig.~\ref{fig.AKT-CA} the difference between \texttt{EVENT2}
distributions in anti--$\kt$ and C--A algorithms, $(\d\sigma^{\akt}_{e}/\d L -
\d\sigma^{\ca}_{e}/\d L)/\so$, for all colour pieces. The slopes for the
$\CF^{2}$ and $\CF\CA$ indicate that an NLL positive $\Rs$--dependent term, and possibly
an NNLL term as well, have been induced by clustering. This is confirmed
in table~\ref{tab:H21_akt}. Moreover, the
fact that the difference between the latter distributions in the $\CF\TF\nf$
piece is non--vanishing implies an NNLL impact of clustering. 
  
Furthermore, we note from fig.~\ref{fig.CA_CF2_zoom} that $C_{2}^{P}$ seems to
slightly vary with the jet--radius parameter $\Rs$. This can be seen for large
values of $\Rs$ ($\Rs = 0.3$) where our small angles
approximation~\eqref{clust_CA_CF2_theta} is not expected to apply.
\begin{figure}[t]
	\centering
	\includegraphics[width=7.5cm]{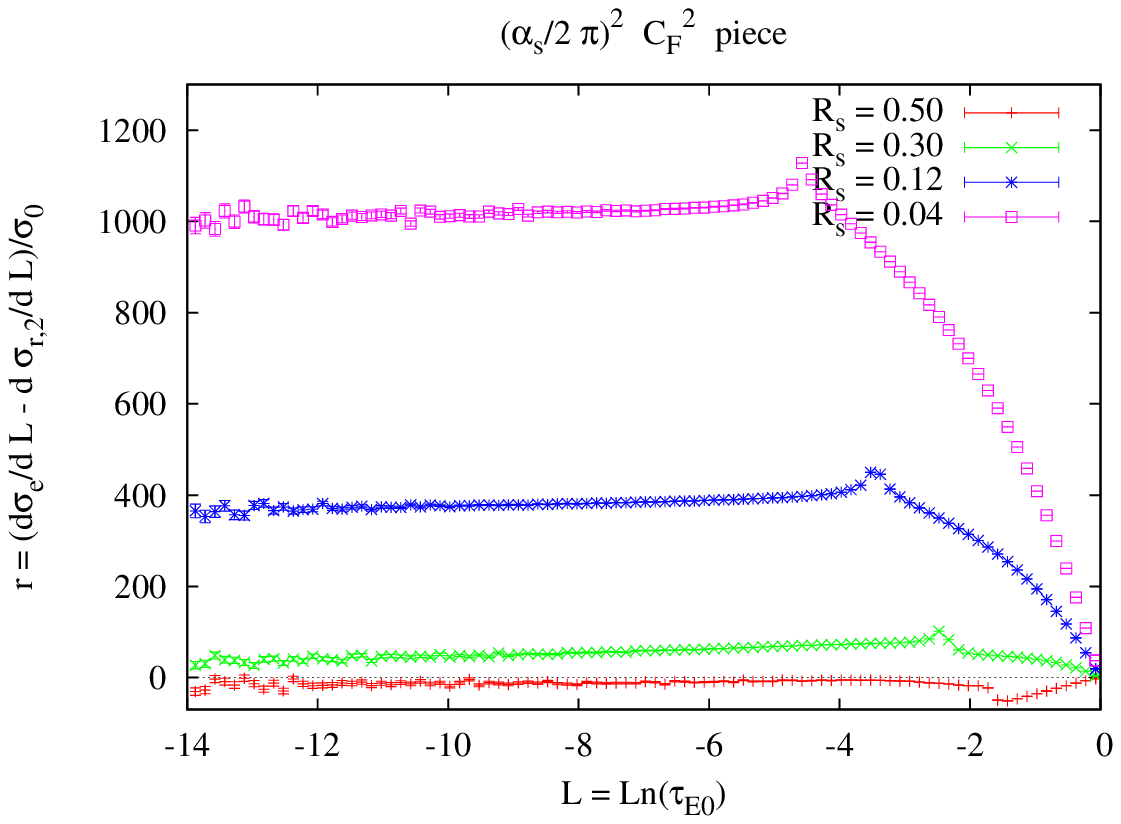}
        \includegraphics[width=7.5cm]{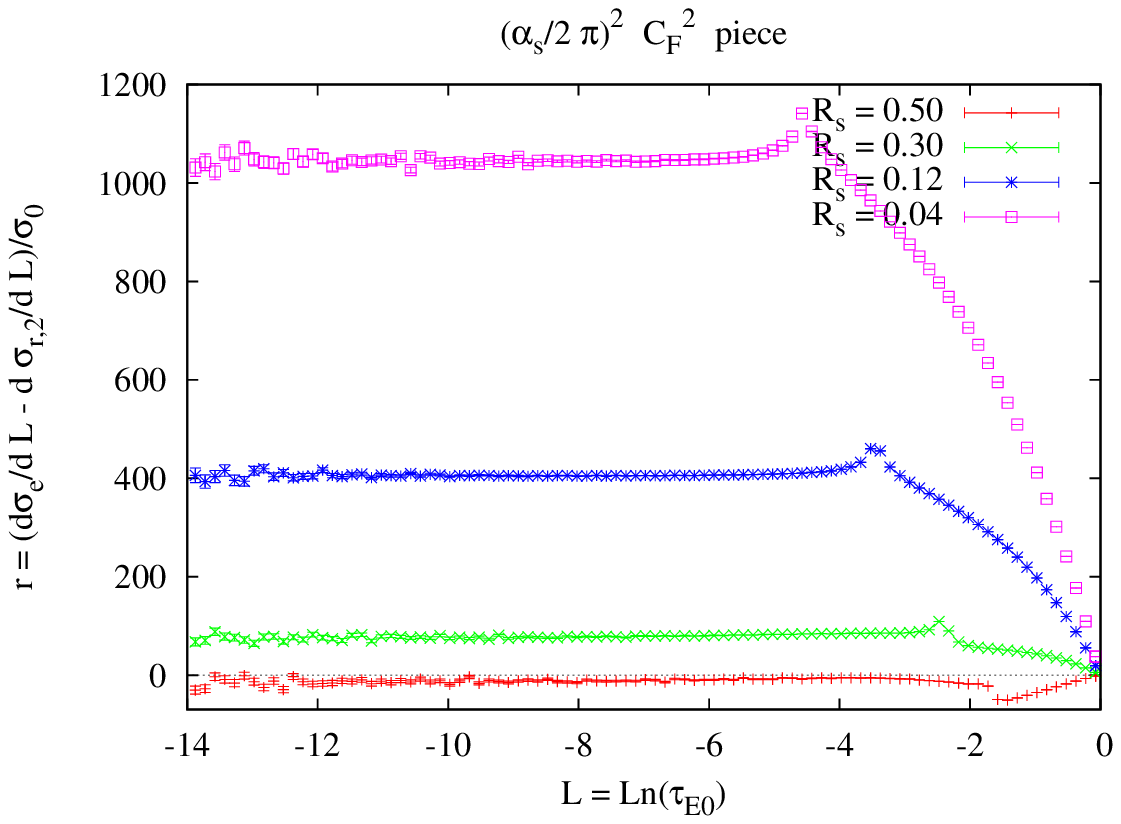}
	\caption{The $\CF^{2}$ part of the difference between \texttt{EVENT2}
and (left) $\tauo$ primary (global) distribution and (right) $\tauo$
distribution including CLs for various jet radii in the C--A algorithm.}
	\label{fig.CA_CF2}
\end{figure}
\begin{figure}[t]
	\centering
	\includegraphics[width=7.5cm]{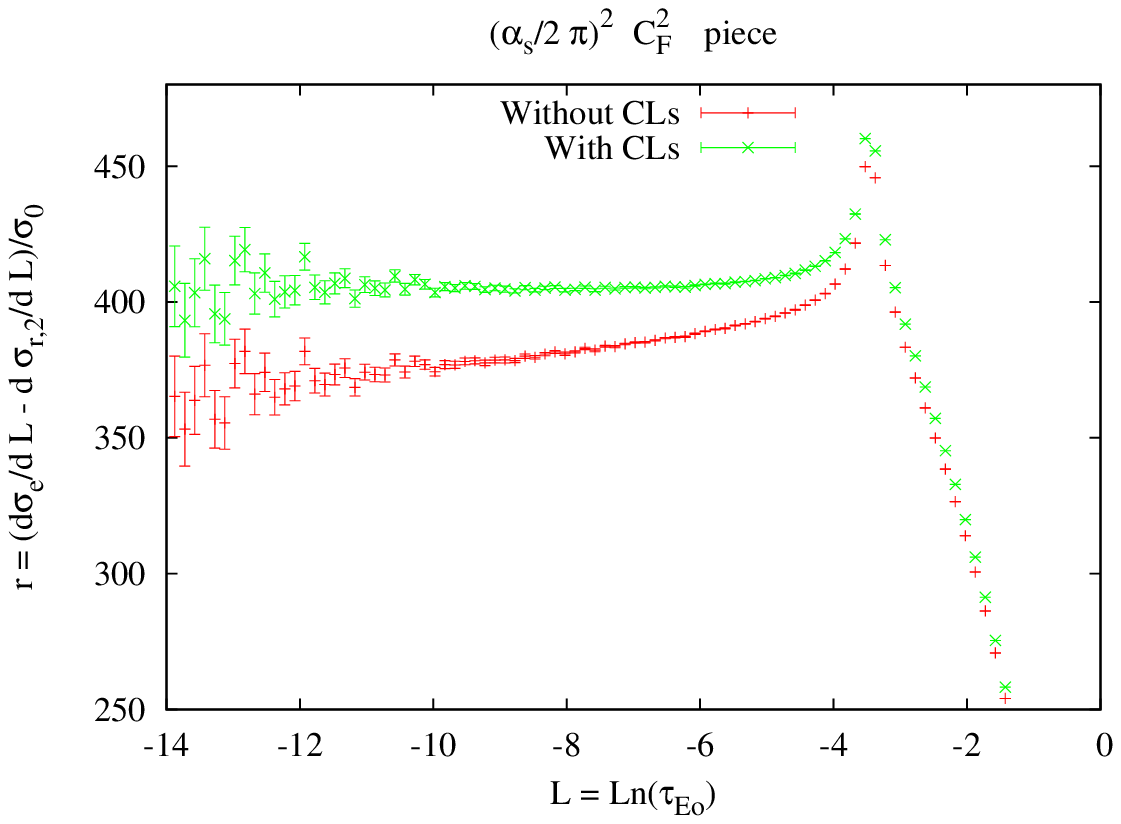}
	\includegraphics[width=7.5cm]{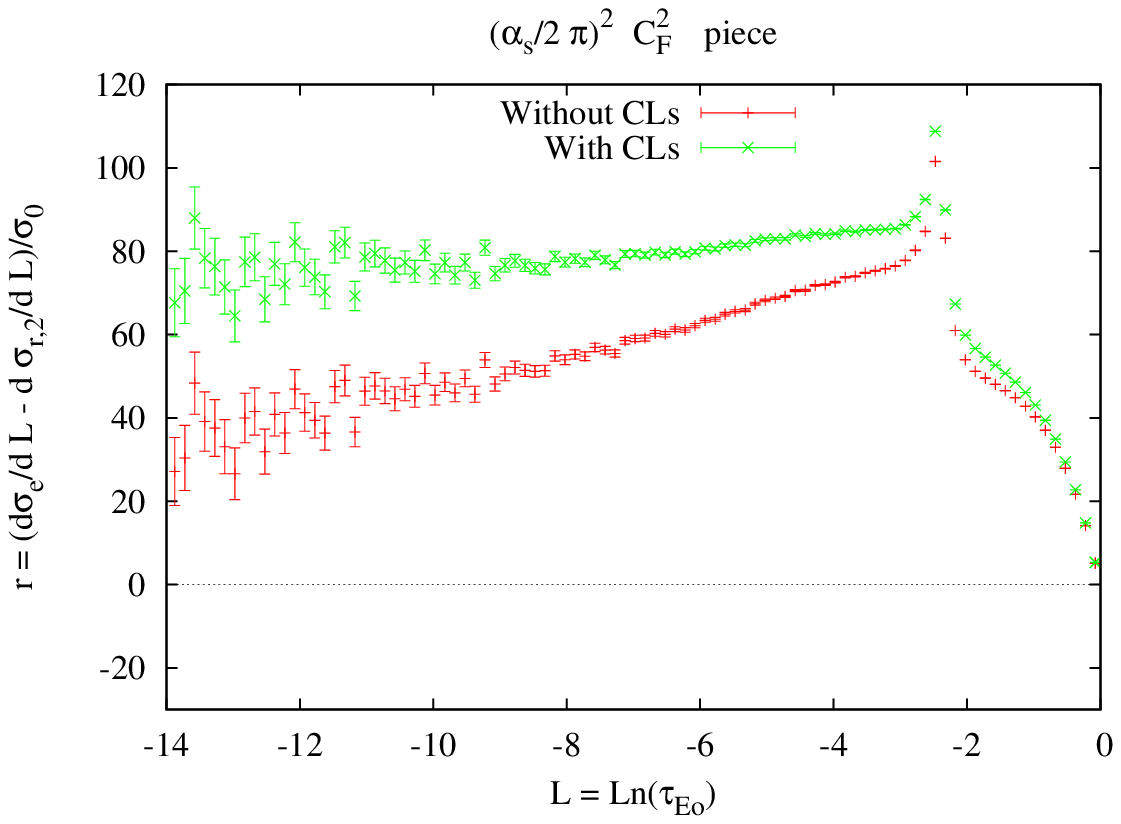}
	\caption{Zoomed--in plots for the $\CF^{2}$ part of the difference
between \texttt{EVENT2} and analytical $\tauo$ distribution with and without CLs
for (left) $\Rs=0.12$ and (right) $\Rs=0.3$ in the C--A algorithm.}
	\label{fig.CA_CF2_zoom}
\end{figure}
  \begin{figure}[t]
	\centering
	\includegraphics[width=7.5cm]{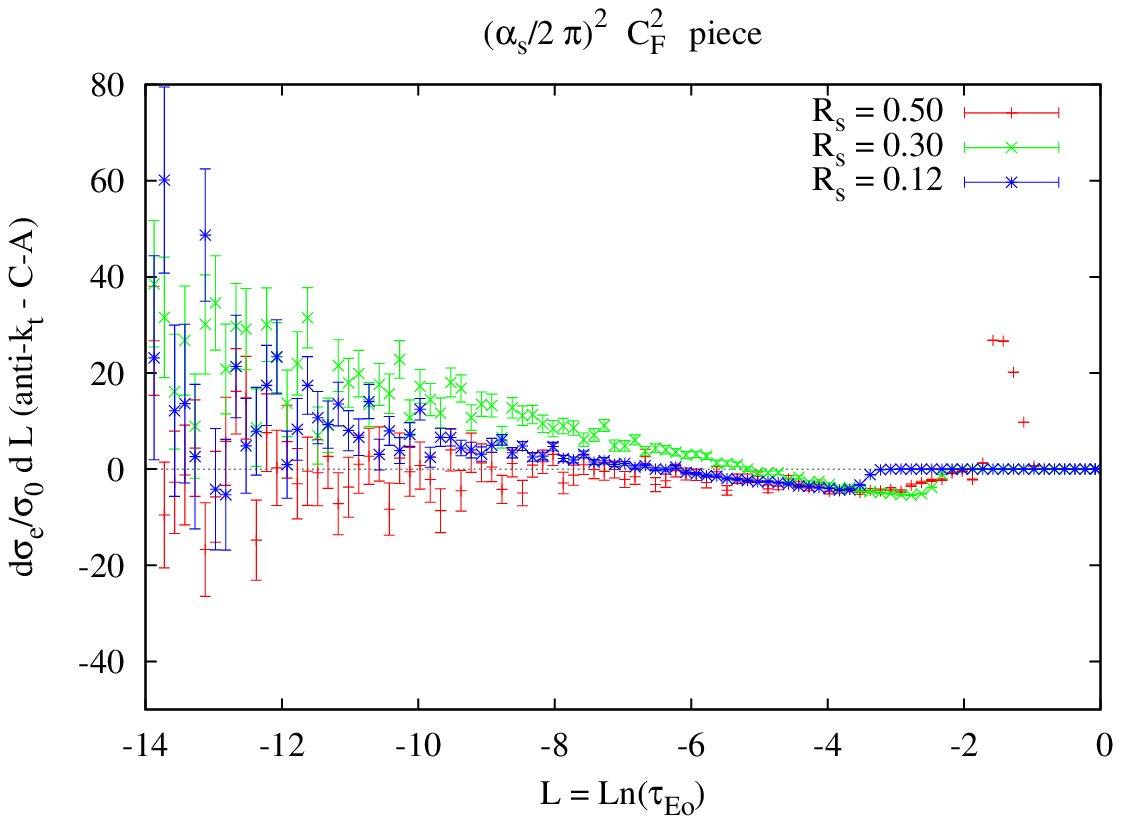}
        \includegraphics[width=7.5cm]{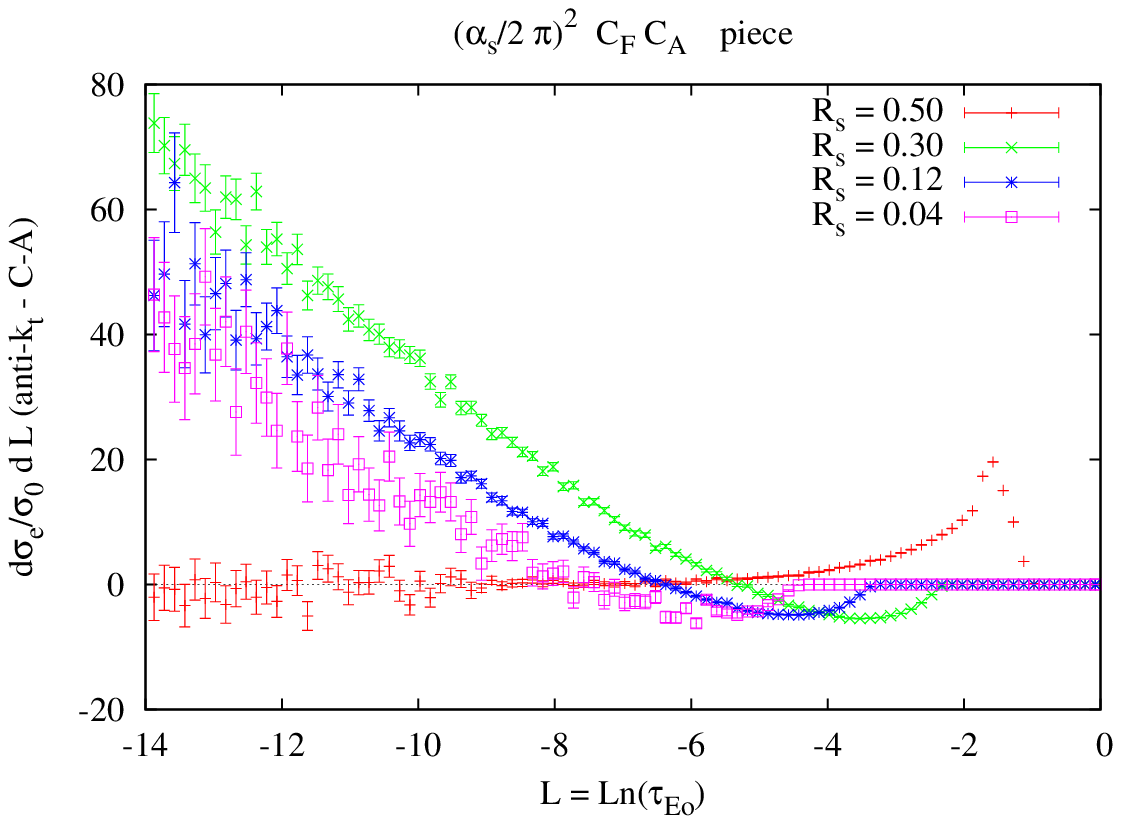}
        \includegraphics[width=7.5cm]{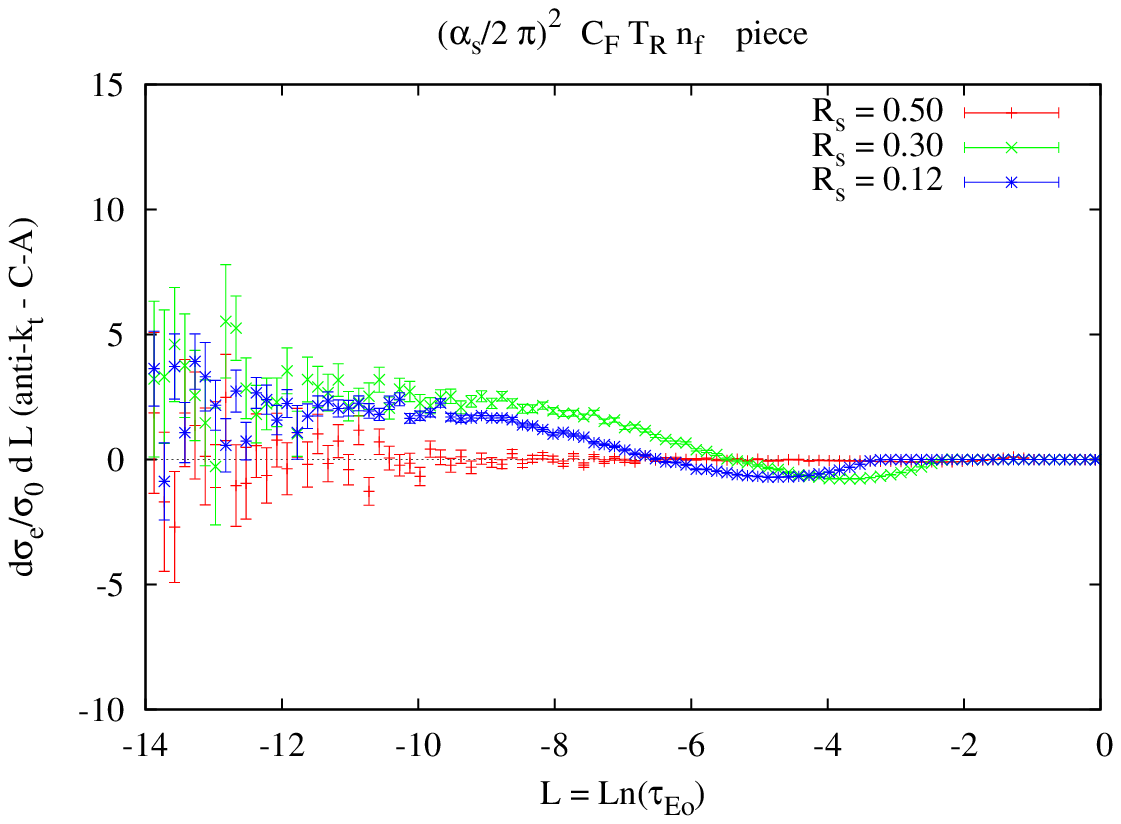}
	\caption{Plots of the three colour pieces of the difference between two
\texttt{EVENT2} distributions corresponding to anti--$\kt$ and C--A algorithms
for various jet radii. We only show $\CF^{2}$ and $\CF\TF\nf$ results at three
values of the jet--radius due to large errors in these colour channels.}
	\label{fig.AKT-CA}
\end{figure}
  \item Similar analysis to those carried in the anti--$\kt$ algorithm apply to
the $\CF\CA$ piece of the $\tauo$ distribution in the C--A algorithm. Including
the NGLs makes the $r(L)$ curve looks convincingly flat in the region $L
\lesssim -9$, particularly for smaller values of $\Rs$, as shown in
fig.~\ref{fig.CA_CFCA}. Recall that we have used the small $\Rs$ limit in
carrying out the computation of $S_{2}^{\ca}$, eq.~\eqref{S2_CA_theta}. Our findings agree with those reported in~\cite{Hornig:2011tg} for jet--radii up to $\Rs \sim 0.3\;(R \sim 1)$.

For completeness, the $\CF\TF\nf$ piece of the $r(L)$ in the C--A algorithm is
depicted in fig.~\ref{fig.CA_CFTF}. As shown in table~\ref{tab:H21_akt}, clustering requirement again reduces the impact of the subleading NGLs in both $\CF\CA$ and $\CF\TF\nf$ channels.
\begin{figure}[t]
	\centering
	\includegraphics[width=7.5cm]{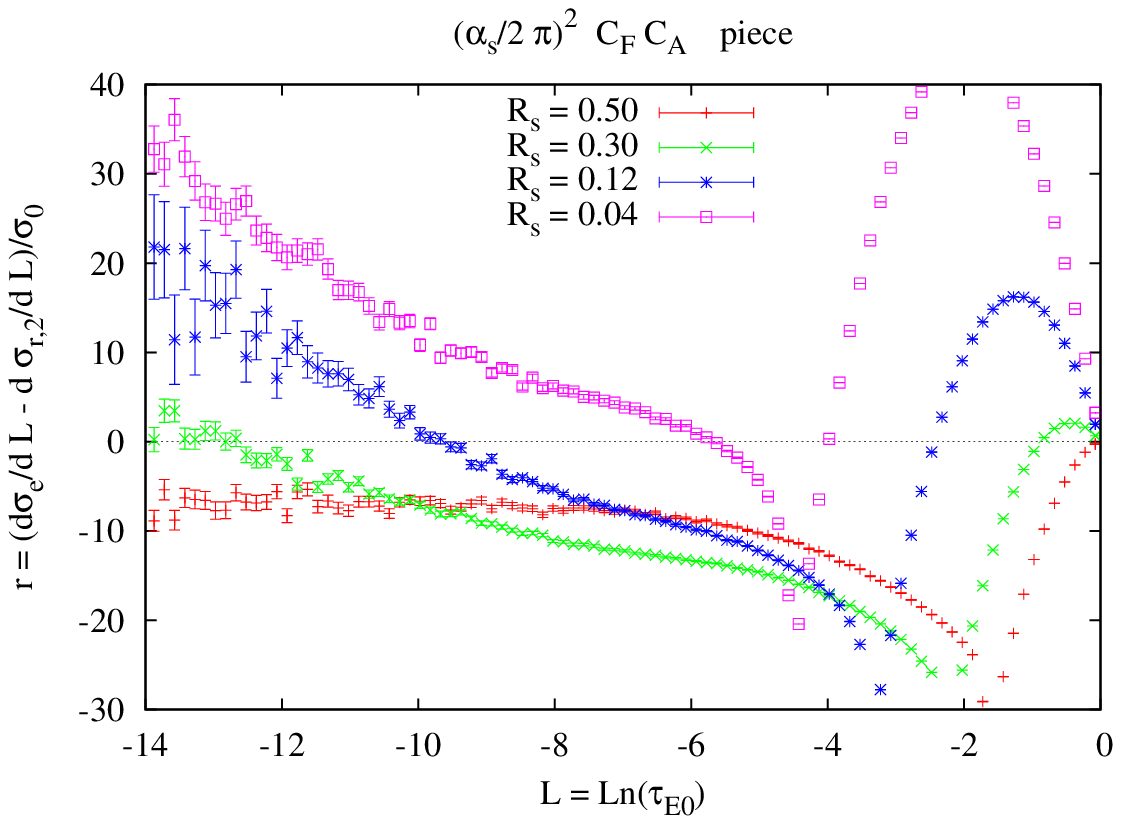}
        \includegraphics[width=7.5cm]{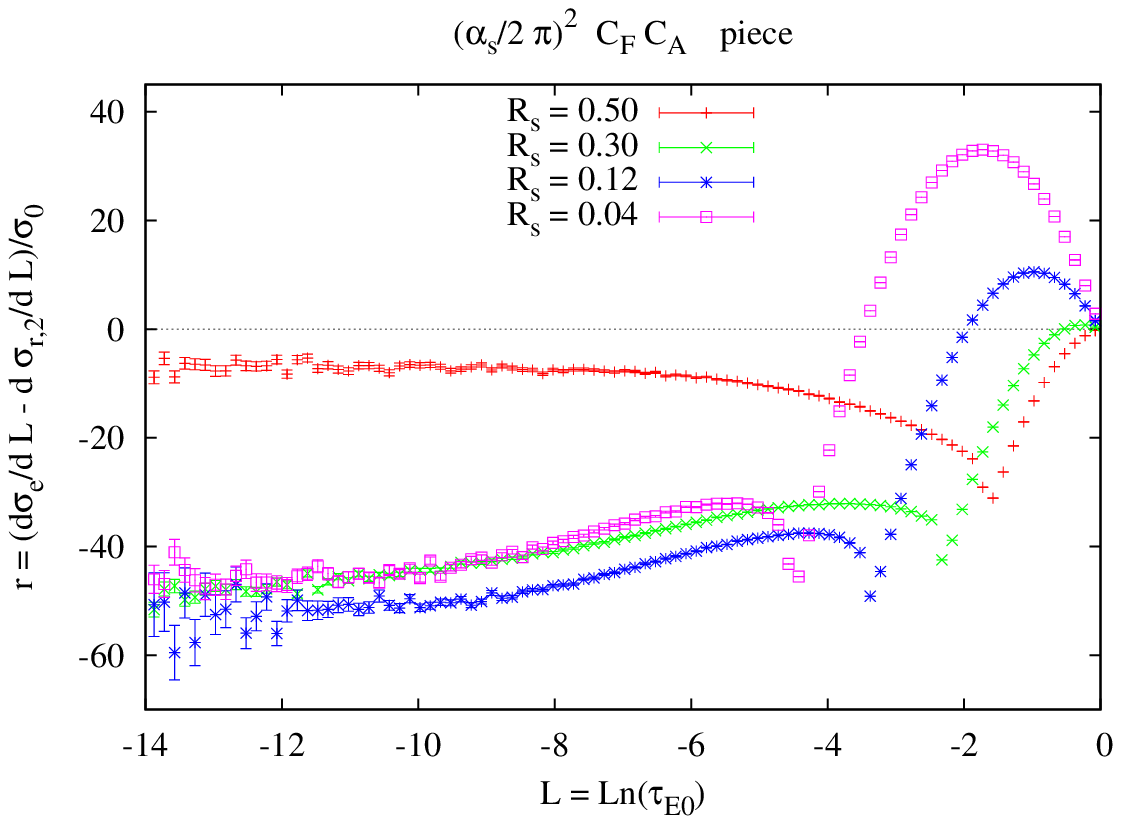}
	\caption{The $\CF\CA$ part of the difference between \texttt{EVENT2} and
(left) $\tauo$ primary (global) distribution and (right) $\tauo$ distribution
including NGLs for various jet radii in the C--A algorithm.}
	\label{fig.CA_CFCA}
\end{figure}
\begin{figure}[t]
	\centering
	\includegraphics[width=7.5cm]{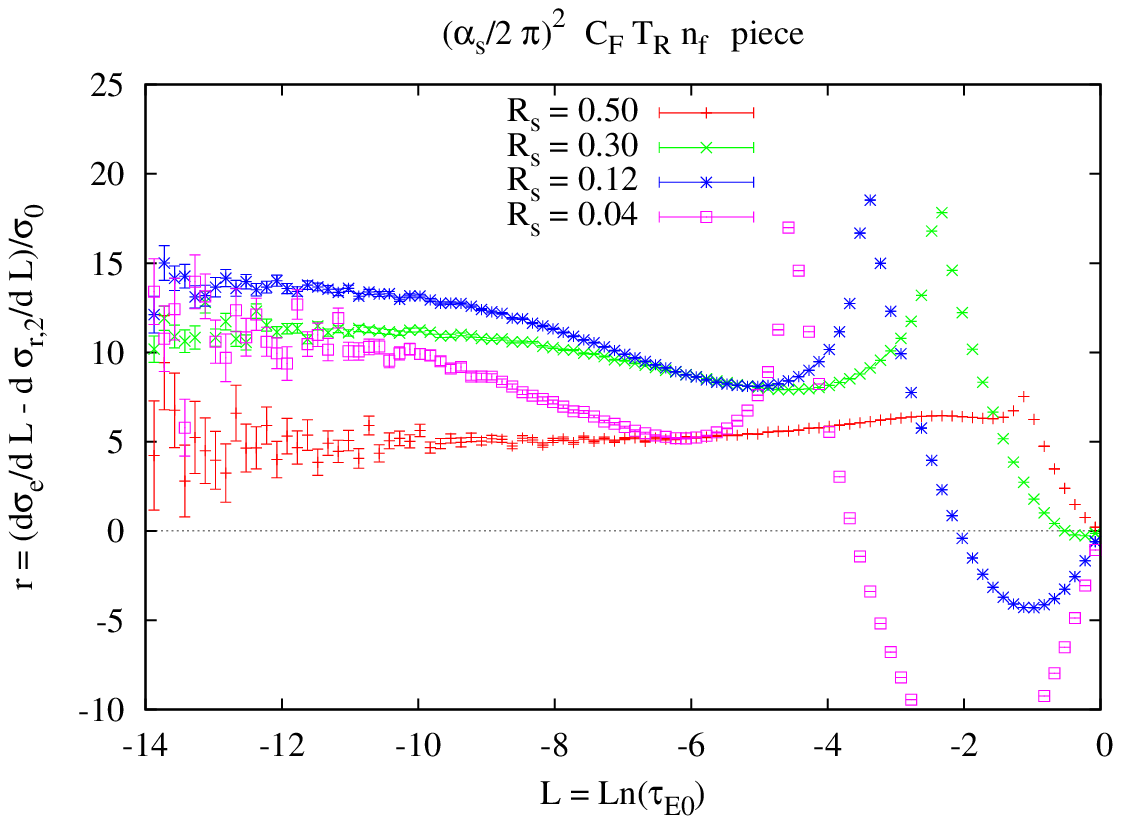}
	\caption{The $\CF\TF\nf$ piece of the difference between \texttt{EVENT2}
and $\tauo$ distribution for various jet radii in $\ca$ algorithm.}
	\label{fig.CA_CFTF}
\end{figure}
\end{itemize}
In summary, we  have confirmed through explicit comparison to exact numerical
distributions the existence of large NGLs and large CLs for the $\tauo$
distribution at NLL and beyond. In light of these findings, the surprising cancellation between
primary--only analytical distribution and \texttt{EVENT2} presented in~\cite{Kelley:2011tj} v$1$ may be explained as follows. While each colour part ($\CF^{2}, \CF\CA$ and $\CF\TF\nf$) separately does not agree with \texttt{EVENT2}, as shown in figs.~\ref{fig.CA_CF2},~\ref{fig.CA_CF2_zoom},~\ref{fig.CA_CFCA} and~\ref{fig.CA_CFTF}, their sum seems to agree with \texttt{EVENT2} (recall that in  of~$[4]$ v$1$ only the sum of the three colour factors is plotted against \texttt{EVENT2}). Such an unexpected agreement can arise from the following possible sources:
\begin{itemize}
     \item The $\ln(\tauo)$ region considered in~\cite{Kelley:2011tj} v$1$ does not correspond to the asymptotic region where large logs are expected to dominate over non--logarithmic terms. Thus the agreement shown in plot~$1$ of~\cite{Kelley:2011tj} v$1$ does not convey any message and all one can say is that the non--logarithmic terms in the range $[-9,0]$ happen to cancel out (see fig~\ref{fig.CA_tot}).    
    \item NGLs are significantly reduced in the $\ca$ algorithm especially for a jet radius $\Rs =0.3$ (which is the one considered in~\cite{Kelley:2011tj} v$1$), as clearly seen in figs.~\ref{fig.NG_coff_CA} and~\ref{fig.CA_CFCA} (left). For smaller jet radii, there is a clear disagreement between the result of~\cite{Kelley:2011tj} v$1$ and \texttt{EVENT2} as shown in fig.~\ref{fig.CA_tot}
     \begin{figure}[h]
      \centering
      \includegraphics[width=12cm]{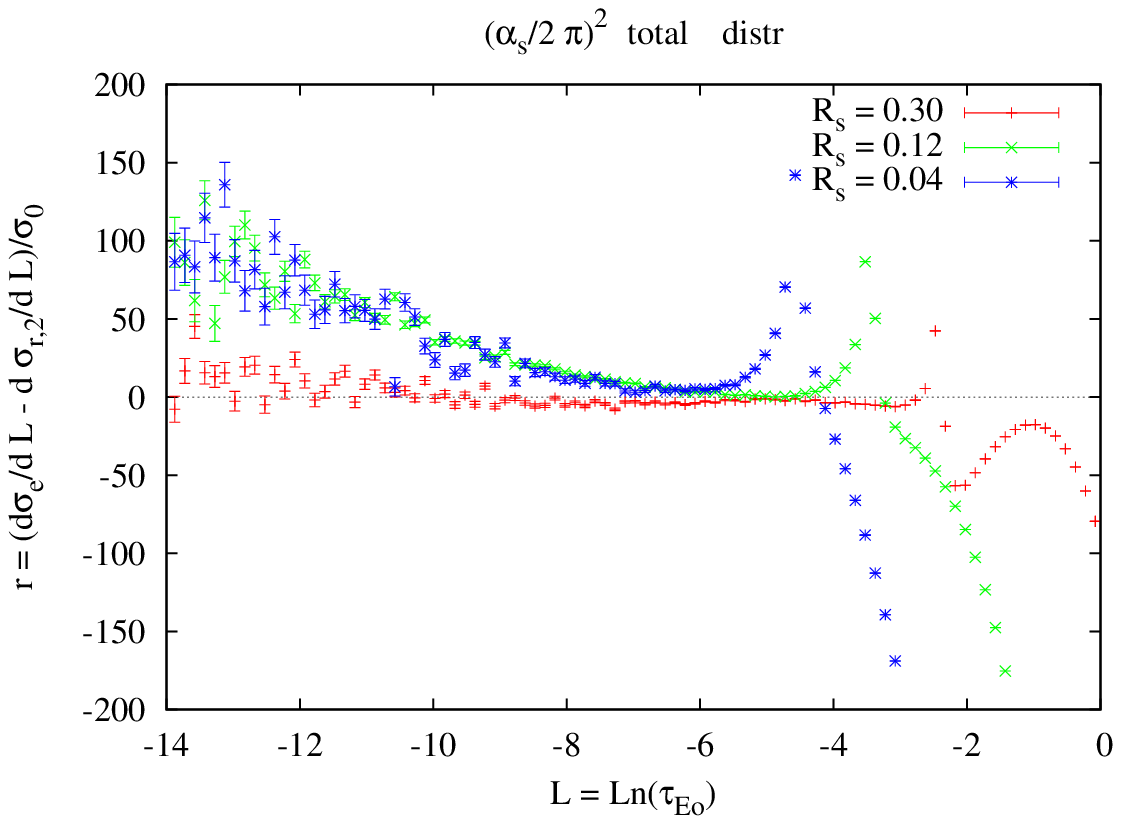}  
      \caption{Difference between the sum of the three colours and \texttt{EVENT2} for various jet radii.}
      \label{fig.CA_tot} 
     \end{figure}
\end{itemize}
We have also shown that clustering the final state partons with the C--A algorithm yielded a significant reduction in NGLs impact, at NLL and beyond, albeit inducing large CLs, at NLL and beyond, in the primary emission sector.

\section{Conclusion}\label{sec.conclusion}

The jet mass with a jet veto, or simply the threshold thrust, is an example of a
wider class of non--global observables. These have the characteristic of being
sensitive to radiation into restricted regions of phase space, or sensitive to
radiation into the whole phase space but differently in different regions. For
such observables the universal Sudakov form factor fails to reproduce the full
logarithmic structure even at NLL accuracy. New contributions that are dependent on
various variables such as the jet size and jet definition
appear at this logarithmic level. In this paper, we have elaborated on these very
contributions for the aforementioned observable. 

Considering secondary emissions, we have computed the full analytical expression
of the first term, $S_{2}$, in a series of missing large logs, namely NGLs. The
coefficient depends, as anticipated, on the jet size and saturates at its
maximum in the limit where the latter, i.e, jet size, vanishes. This saturation
value was used in~\cite{Banfi:2010pa} as an approximation to the full value in
the small $\Rs$ limit. It turns out that the approximation is valid for quite a
wide range of $\Rs$. The formula for $S_{2}$ has been checked against full exact
numerical result obtained by the program \texttt{EVENT2}. The difference between
the analytical and numerical differential distributions was shown to be
asymptotically flat signalling a complete cancellation of singular terms up to
NLL level.  This has all been done for final states defined in the cone--like
anti--$\kt$ jet algorithm.

To illustrate the dependence of NLL on the jet definition, we have
investigated the effects of applying the C--A algorithm on $\ee$ final states.
The impact of soft partons clustering is two--fold. On one side, it reduces the
size of NGLs through shrinking the phase space region where the latter
dominantly come from. i.e, the region where the emitter and emitted soft partons
are just in and just out of the jet. On the other side, it gives rise to new NLL
logarithmic contributions, CLs, in the primary emission sector. In the small
jet--radius limit, the corresponding coefficient at second order has been shown, through comparison to \texttt{EVENT2}, to be independent of $\Rs$.
 
Furthermore, our numerical analyses with \texttt{EVENT2} have shown that the asymptotic region where the said large logs, in both anti--$\kt$ and C--A jet algorithms, dominate
corresponds to $L \lesssim -9$ and decreases for smaller values of the
jet--radius. As a by--product, we have found that there are subleading NGLs in both $\CF\CA$ and $\CF\TF\nf$ pieces as well as subleading CLs in the $\CF^{2}$ piece of the $\tauo$ distribution. Clustering impact on NGLs has been observed to extend to NNLL level too. Regarding NGLs in $\CF\CA$ channel in both jet algorithms, our findings serve as a confirmation of the corresponding calculations performed within SCET in~\cite{Hornig:2011tg}.

Based on our rough approximation to NLL resummation, which is exponentiating the fixed--order result for both NGLs and CLs, it has been shown that it may be possible to completely eliminate the non--global correction to the primary Sudakov form factor at all--orders for events where final states clustering is applied. This elimination can be achieved by tuning the jet--radius parameter, $\Rs$, of the jet algorithm as well as the jet veto $\Eo$. If such \emph{optimal} values of $\Rs$ and $\Eo$ are of practical significance, that is $ R \sim 0.4$ or so and $\Eo \gg \Lambda_{\mathrm{QCD}}$, then the single--gluon exponentiation should be sufficient in describing the experimental data. A concrete answer of whether such optimal values exist can only be established once an all--orders resummation of primary, NGLs and CLs is performed. We postpone this investigation to future publications.

As mentioned earlier in sec.~\ref{sec.fixed_order_2} and shown
in~\cite{Banfi:2010pa}, the inclusive $\kt$ jet algorithm behaves in an
identical way to C--A algorithm with regard to the threshold thrust
distribution. It would be interesting to conduct similar studies for events
defined in IRC cone algorithms such as the SISCone. In
principle, one expects to see analogous effects not only for the threshold
thrust but for all shape variables that are of non--global nature. Moreover, we reserve
the extension of the findings of this paper to hadron--hadron collisions to
future work. Apart from complications due to coloured initial state, we expect
the gross features of this paper to apply.

\subsection*{Acknowledgement}
 
I am indebted to M.~Dasgupta, S.~Marzani and A.~Banfi for collaboration on
related work and helpful discussions on the current paper. I would like to thank
M.~Seymour for his generous help and useful feedback.

\appendix

\section{Derivation of LO distribution}\label{app.LO_distr}

In the present section we outline the derivation of the full logarithmic part of
the LO $\tauo$ integrated distribution~\eqref{R1_full-b}. For the emission of a
single gluon, i.e, $\ee \rightarrow q\,\qbar\,g$, we define the kinematic
variables, $x_{i} = 2p_{i}.Q/Q^{2} = 2 E_{i}/Q$ and $y_{ij} = 2p_{i}.p_{j}/Q^{2}
= 1-x_{k}$ where $i,j,k = 1(q), 2(\qbar), 3(g)$. The $\Or(\as)$ matrix--element
squared can be computed by considering two Feynman graphs corresponding to real
emission of the gluon $g$ off the two hard legs $q, \qbar$. Applying the
appropriate QCD Feynman rules and supplementing the three--body phase space
factor, the corresponding differential distribution is given by
\begin{equation}\label{dsig1_sig}
\frac{\d^{2}\sigma^{(1)}}{\sigma \d x_{1} \d x_{2}}= \frac{\CF \as}{2\pi}
\frac{x_{1}^{2} + x_{2}^{2}}{(1-x_{1})(1-x_{2})},
\end{equation} 
where $\sigma$ is the total hadronic cross--section. Up to $\Or(\as^{2})$, it is
given in terms of the Born cross--section, $\so$, by the
relation~\cite{Appelquist:1973uz}
\begin{equation}\label{sig_had-sig_0}
\frac{\sigma}{\sigma_{0}} = 1 + \frac{\as}{2\pi} \left[\frac{3\CF}{2}\right] +
\left(\frac{\as}{2\pi}\right)^{2} K_{2} + \Or(\as^{3}).
\end{equation}
with
\begin{equation}\label{sig_sig0_K}
K_{2} = -\CF^{2} \frac{3}{8} + \CF\CA\left(\frac{123}{8} -11\zeta_{3}\right)
+\CF\TF\nf \left(-\frac{11}{2} + 4\zeta_{3}\right).
\end{equation}
The integration region, which is originally $1\geq x_{1},x_{2} \geq 0$ and
$x_{1}+x_{2} \geq 1$ and which leads to divergences, gets modified by
introducing the jet shape variable. For three partons in the final state,
$\tauo$ is zero unless two partons are clustered together. Therefore $\tauo$ is
non--vanishing only in two--jet events. For the latter events, there are six
ways of ordering the energy fractions $x_{i}$ corresponding to six regions of
phase space that needs to be integrated over. Due to $x_{1} \leftrightarrow
x_{2}$ symmetry of the matrix--element~\eqref{dsig1_sig}, one can only consider
three regions and multiply the result by a factor of $2$. These regions
correspond to; $x_{1}>x_{2}>x_{3}, x_{1}>x_{3}>x_{2}$ and $x_{3}>x_{1}>x_{2}$.
The threshold thrust is then given by
\begin{multline}\label{tauo_LO}
 \tauo = (1- x_{1}) \Theta(x_{1}-x_{2},x_{2}-x_{3})\Theta\left(2\Rs -1+\cos\theta_{23} \right) +
 \\
  + (1 - x_{1}) \Theta(x_{1}-x_{3},x_{3}-x_{2})\Theta\left(2\Rs-1+ \cos\theta_{23} \right) +
\\
  + (1-x_{3}) \Theta(x_{3}-x_{1}, x_{1}-x_{2}) \Theta\left(2\Rs-1+\cos\theta_{12} \right),
\end{multline}
where $\Theta(a-b,b-c) = \Theta(a-b) \Theta(b-c)$.
To obtain the full logarithmic contribution it is sufficient to only consider
regions where the gluon is the softest parton ($x_{3} = \min(x_{i})$). Other
regions, last term in RHS of eq.~\eqref{tauo_LO}, only contributes
non--logarithmically. Adding up real and virtual contributions,
in~\eqref{S1_tauo_dist}, one is only left with the virtual corrections in the
range $\Theta(1- x_{1} - \tauo)$. The corresponding angular function
in~\eqref{tauo_LO} may be written in terms of the energy fractions as, 
\begin{equation}\label{clust_cond_LO}
1- \cos\theta_{23} = \frac{2(1-x_{1})}{x_{2} x_{3}} \approx
\frac{2(1-x_{1})}{x_{3}},
\end{equation}
where the last approximation follows from the fact that the gluon is the
softest, $ x_{1}, x_{2} \gg x_{3}$. Hence the two--jet contribution to the first
order shape fraction $\Sigma^{(1)}$ is given by
\begin{equation}\label{Sig1_2-jet}
\Sigma^{(1)}(\tauo, \Eo) = -\frac{\CF\,\as}{2\pi}
\int_{1-\Rs(1-\tauo)}^{1-\tauo}\d x_{2} \int_{1+\tauo+x_{2}}^{\frac{x_{2}
-1+\Rs(2-x_{2})}{\Rs}}\d x_{1}\,\frac{x_{1}^{2} + x_{2}^{2}}{(1-x_{1})
(1-x_{2})} \Theta\left(\frac{\Rs}{1+\Rs} - \tauo\right)
\end{equation}

In case of events with three--jets in the final state, the energy of the softest
jet is vetoed to be less than $\Eo$. The corresponding phase space constraint,
left after real--virtual mis--cancellation, on the differential
cross--section~\eqref{dsig1_sig} reads
\begin{equation}\label{E0_constr_LO}
- \Theta\left(x_{3} -\frac{2\Eo}{Q}\right) \Theta\left(\frac{1-x_{1}}{x_{3}} -
\Rs\right)\,\Theta\left(1-\Rs - \frac{1-x_{1}}{x_{3}}\right).
\end{equation}
Noting that $x_{1} + x_{2} + x_{3} = 2$, one can obtain the corresponding
integration limits on $x_{1}$ and $x_{2}$. Adding up the result of the latter
integration with that of eq.~\eqref{Sig1_2-jet} and making use of
the following dilogarithm identities~\cite{polylogs}
\begin{eqnarray}\label{polylogs_rel}
\nonumber \Li_{2}(x) + \Li_{2}(1-x) &=& \frac{\pi^{2}}{6} - \,\ln(x)
\,\ln(1-x),\\
\Li_{2}(x) + \Li_{2}\left(\frac{1}{x}\right) &=& \frac{\pi^{2}}{3} -\frac{1}{2}
\,\ln^{2}(x).
\end{eqnarray}
one obtains eq.~\eqref{R1_full-b}.

\section{$G_{nm}$ coefficients}\label{app.coeff_in_expansion}

The resultant coefficients from the expansion of the exponent in the resummed
integrated distribution, eq.~\eqref{resum-form_QCD-b}, are
\begin{eqnarray}\label{G_nm-QCD}
\nonumber G_{12} &=& -2 \CF ,
\\
\nonumber G_{11} &=& \CF \left(3 - 4 L_{\Rs}\right),
\\
\nonumber G_{10} &=& \CF\left[-4 L_{\Rs} L_{\Eo} +
\frac{\bar{f}_{0}(\Rs)}{2}\right],
\\
\nonumber G_{23} &=&  \CF \left(\frac{4}{3} \TF \nf - \frac{11}{3}\CA \right),
\\
\nonumber G_{22} &=& -\frac{4 \pi^{2}}{3} \CF^{2} + \CF
\CA\left(\frac{\pi^{2}}{3} - S_{0}(\Rs) -\frac{169}{36} - \frac{22}{3}
L_{\Rs}\right) + \CF \TF\nf \left(\frac{11}{9} + \frac{8}{3} L_{\Rs}\right),
\\
\nonumber G_{21} &=& - \CF^{2}\,\frac{8 \pi^{2}}{3} L_{\Rs} - \CF\CA\bigg[2
S_{0}(\Rs) L_{\Eo} - \left(\frac{2\pi^{2}}{3} -2S_{0}(Rs) - \frac{134}{9} -
\frac{11}{3} L_{\Rs}\right) L_{\Rs}\bigg] +\\
&+& \CF\TF\nf \left(\frac{4}{3} L_{\Rs} + \frac{40}{9}\right) L_{\Rs}.
\end{eqnarray}
where $L_{\Rs} = \ln(\Rs/(1-\Rs))$ and $L_{\Eo} = \ln(2\Eo/Q)$. The factor
$\bar{f}_{0}(\Rs)$ only captures the first term of $f_{0}$ given in
eq.~\eqref{f_omeg}. We simply replace $\bar{f}_{0} \mapsto f_{0}$ when comparing
to the numerical distribution. Moreover, we have introduced, for shorthand, the
function $S_{0}(\Rs)$ given by (cf. eq.~\eqref{S2_akt_a1}),
\begin{equation}\label{S_0_Rs}
S_{2} = -\CF\CA\; S_{0}(\Rs).
\end{equation}
The one--loop constant is given by, eq.~\eqref{R1_full-b},
\begin{eqnarray}\label{C1_QCD}
C_{1} &=& \CF \left(-1+\frac{\pi^{2}}{3}\right),
\end{eqnarray}
Expanding the total resummed distribution in eq.~\eqref{resum-form_QCD-b} to
$\Or(\as^{2})$ and up to NLL we have
\begin{multline}\label{resum_tot_series}
\Sigma_{r,2}(\widetilde{L}) = 1 + \left(\frac{\as}{2\pi}\right)\left(H_{12}
\widetilde{L}^{2} + H_{11} \widetilde{L} + H_{10}\right) + 
\left(\frac{\as}{2\pi}\right)^{2} \Big(H_{24} \widetilde{L}^{4} + H_{23}
\widetilde{L}^{3} + H_{22} \widetilde{L}^{2} + \\+ H_{21} \widetilde{L} +
H_{20}\Big),
\end{multline}
where (recall that $\widetilde{L} = \ln(1/\tauo) \Rightarrow \tauo =
e^{-\widetilde{L}}$)
\begin{eqnarray}\label{H_nm_coeffs}
\nonumber D_{\mathrm{fin}} (e^{-\widetilde{L}}) &=&
\left(\frac{\as}{2\pi}\right)\,d_{1}(e^{-\widetilde{L}}) +
\left(\frac{\as}{2\pi}\right)^{2}\,d_{2}(e^{-\widetilde{L}}),\\ 
\nonumber H_{12} &=& G_{12},\\
\nonumber H_{11} &=& G_{11},\\
\nonumber H_{10} &=& G_{10} + C_{1} + d_{1}(\tauo),\\
\nonumber H_{24} &=& \frac{1}{2} G_{12}^{2}.\\
\nonumber H_{23} &=& G_{23} + G_{12} G_{11},\\
\nonumber H_{22} &=& G_{22} + (G_{10} + C_{1}) G_{12} + \frac{1}{2}
G_{11}^{2},\\
\nonumber H_{21} &=& G_{21} + (G_{10}+ C_{1}) G_{11},\\
H_{20} &=& G_{20} + \frac{1}{2} G_{10}^{2} + C_{1} G_{10} + C_{2} +
d_{2}(\tauo). 
\end{eqnarray}
Differentiating~\eqref{resum_tot_series} w.r.t. $\widetilde{L}$, the NLO
differential distribution reads
\begin{equation}\label{resum_expanded-sig-diff}
\frac{\d\Sigma_{r,2}}{\d \widetilde{L}} =
\frac{1}{\sigma_{0}}\frac{\d\sigma_{r,2}}{\d \widetilde{L}} =
\delta(\widetilde{L})\, D_{\delta} + \left(\frac{\as}{2\pi}\right)\,
D_{A}(\widetilde{L}) + \left(\frac{\as}{2\pi}\right)^{2}\, D_{B}(\widetilde{L}),
\end{equation}
where the singular (logarithmic) terms are given by
\begin{eqnarray}\label{singular_terms_gen}
\nonumber D_{\delta} &=& 1 + \left(\frac{\as}{2\pi}\right) \left[G_{10} +
C_{1}\right] + \left(\frac{\as}{2\pi}\right)^{2} \left[G_{20} + \frac{1}{2}
G_{10}^{2} + C_{1} G_{10} + C_{2}\right],
\\
\nonumber D_{A}(\widetilde{L}) &=& 2 H_{12} \widetilde{L} + H_{11} +
\frac{\d}{\d \widetilde{L}}\, d_{1}(e^{-\widetilde{L}}),
\\
 D_{B}(\widetilde{L}) &=& 4 H_{24} \widetilde{L}^{3} + 3 H_{23}
\widetilde{L}^{2} + 2 H_{22} \widetilde{L} + H_{21} + \frac{\d}{\d
\widetilde{L}}\, d_{2}(e^{-\widetilde{L}}).
\end{eqnarray}

\section{Threshold thrust distribution in SCET}\label{app.tw_in_SCET}

The resummation of the threshold thrust in SCET is presented in the current
section for comparison with pQCD. We shall only present the final form of the
resummed result taken from
Refs.~\cite{Kelley:2011tj,Kelley:2010qs,Becher:2008cf}. For a full derivation
and more in depth discussion one should consult the latter references. The only
task we have performed here is the expansion of the full resummed distribution
to $\Or(\as^{2})$.

\subsection{Resummation}\label{subsec.resummation_SCET}

The general formula of the resummed distribution for the threshold thrust is
given by~\cite{Kelley:2011tj,Becher:2008cf}
\begin{multline}\label{resum-form_SCET-a}
\frac{\d\Sigma^{\SCET}(\tauo, R)}{\d\tauo} =
\frac{\d\sigma^{\SCET}}{\sigma_{0}\d\tauo} = \exp\left[4S(\mu_{h},\mu_{j}) +
4S(\mu_{s},\mu_{j}) - 4A_{H}(\mu_{h},\mu_{s}) + 4A_{J}(\mu_{j},\mu_{s}) \right]
\\ \times\left(\frac{\Rs}{1- \Rs}\right)^{-2A_{\Gamma}(\mu_{\omega},\mu_{s})}
\left(\frac{Q^{2}}{\mu_{h}^{2}}\right)^{-2A_{\Gamma}(\mu_{h},\mu_{j})}
H(Q^{2},\mu_{h}) \; S^{\outt}_{R}(\omega,\mu_{\omega}) \\ \times
\left[\tilde{j}\left(\ln\frac{\mu_{s}Q}{\mu_{j}^{2}} + \partial_{\eta},
\mu_{j}\right) \right]^{2} \tilde{s}^{\inn}_{\tauo}(\partial_{\eta}, \mu_{s})
\frac{1}{\tauo} \left(\frac{\tauo Q}{\mu_{s}}\right)^{\eta} \frac{e^{-\gamma_{E}
\eta}}{\Gamma(\eta)}.
\end{multline}
See~\cite{Kelley:2011tj,Becher:2008cf} for full notation. In order to compute
the fixed--order expansion of~\eqref{resum-form_SCET-a} up to $\Or(\as^{2})$,
all scales should be set equal ($\mu_{h} = \mu_{j} = \mu_{s} = Q$). In this
limit, the evolution factors $S, A_{J}$ and $A_{H}$ vanish. The differentiation
w.r.t. $\eta$ is carried out using the explicit form of $\tilde{j}$ and
$\tilde{s}^{\inn}_{\tauo}$. The final result of the integrated distribution may
be cast in the generic form~\eqref{resum-form_QCD-b} with the constants and
coefficients of the logs given by
\begin{eqnarray}\label{eq.C1-C2_SCET}
C_{1} &=& \CF\left(-1+\frac{\pi^{2}}{3} \right),
\\
\nonumber C_{2} &=& \CF^{2} \left(1- \frac{3\pi^{2}}{8} +\frac{\pi^{4}}{72}
-6\zeta(3) \right) + \CF\CA \left(\frac{493}{324} + \frac{85\pi^{2}}{24}
-\frac{73\pi^{4}}{360} + \frac{283 \zeta(3)}{18} \right) +
\\
&+& \CF\TF\nf \left(\frac{7}{81} -\frac{7\pi^{2}}{6}
-\frac{22\zeta(3)}{9}\right) + C_{2}^{\inn} + C_{2}^{\outt},
\end{eqnarray}
and
\begin{eqnarray}\label{G_nm_SCET}
\nonumber G_{12} &=& -2 \CF,
\\
\nonumber G_{11} &=& -\CF \left(3- 4 L_{\Rs}\right),
\\
\nonumber G_{10} &=& \CF\left(- 4 L_{\Rs} L_{\Eo} + \frac{f_{0}(\Rs)}{2}\right),
\\
\nonumber G_{23} &=& \CF \left(\frac{11}{3}\CA - \frac{4}{3} \TF \nf \right),
\\
\nonumber G_{22} &=& -\frac{4 \pi^{2}}{3} \CF^{2} + \CF
\CA\left(\frac{\pi^{2}}{3}-\frac{169}{36} - \frac{22}{3} L_{\Rs}\right) + \CF
\TF\nf \left(\frac{11}{9} + \frac{8}{3} L_{\Rs}\right),
\\
\nonumber G_{21} &=& \CF^{2}\left[-\frac{3}{4} - \pi^{2} + 4\zeta(3) +\frac{8
\pi^{2}}{3} L_{\Rs}\right] +\\\
\nonumber &+& \CF\CA\bigg[-\frac{57}{4} + 6\zeta(3) - \left(\frac{2\pi^{2}}{3}
-\frac{134}{9} -\frac{11}{3} L_{\Rs}\right) L_{\Rs}\bigg] +
\\
\nonumber &+& \CF\TF\nf\left[5 - \left(\frac{4}{3} L_{\Rs}+ \frac{40}{9}\right)
L_{\Rs}\right],\label{G21_SCET}
\end{eqnarray}
\begin{eqnarray}\label{G_nm_SCET_2}
\nonumber G_{20} &=& \CF^{2}  \Bigg[-\frac{f_{0}^{2}}{8} + \left(2\pi^{2}
-16\zeta(3)\right) L_{\Rs} - \Big(\frac{11\pi^{2}}{6} + \frac{f_{0}}{2}\Big)
L^{2}_{\Rs} - L_{\Rs}^{2} \Bigg]+
\\
\nonumber &+& \CF\CA \Bigg[\frac{11\pi^{2}}{9} L_{\Rs} - \frac{11}{6} L_{\Rs}
L^{2}_{\Eo} - L_{\Eo} \left(\frac{11f_{0}}{12}
+\left[\frac{134}{9}-\frac{2\pi^{2}}{3}\right] L_{\Rs} +\frac{11}{6}
L^{2}_{\Rs}\right) \Bigg] +
\\
 &+& \CF\TF\nf\Bigg[-\frac{4\pi^{2}}{9} L_{\Rs} + \frac{2}{3} L_{\Rs}
L^{2}_{\Eo} + L_{\Eo}\Bigg(\frac{f_{0}(\Rs)}{3} +\frac{40}{9} L_{\Rs} +
\frac{2}{3} L^{2}_{\Rs}\Bigg) \Bigg].
\end{eqnarray}
Considering primary emission, the only missing piece in the distribution is the
two--loop constants in the soft function, namely $C_{2}^{\inn}$ and
$C_{2}^{\outt}$. 

\bibliographystyle{JHEP3}
\bibliography{Refs}

\end{document}